\documentclass[a4paper,11pt]{article}
\usepackage{float}
\usepackage{jheppub} 
\usepackage{verbatim}

\usepackage[normalem]{ulem}
\usepackage{slashed}
\usepackage{bigints}
\usepackage{braket}
\usepackage{cleveref}
\usepackage{comment}
\usepackage{bm}		    

\def\be{\begin{equation}}
\def\ee{\end{equation}}
\def\figs/B{B}
\def\bea{\begin{eqnarray}}
\def\eea{\end{eqnarray}}
\def\bg{\begin{eqnarray}}
\def\nd{\end{eqnarray}}

\def\log{{\rm log}}

\title{Higgs Inflation: Particle Factory}

\author[a]{Tammi Chowdhury,}
\author[b,c]{Leah Jenks,}
\author[b,d]{Edward W. Kolb,}
\author[e]{Evan McDonough}

\affiliation[a]{Department of Physics \& Astronomy, \\
University of Manitoba, Winnipeg, Manitoba R3T 2N2, Canada}
\affiliation[b]{Kavli Institute for Cosmological Physics, \\
The University of Chicago,  5640 South Ellis Avenue, Chicago, IL 60637, U.S.A.}
\affiliation[c]{William H. Miller III Department of Physics \& Astronomy,\\ Johns Hopkins University, 3400 N. Charles St., Baltimore, MD 21218, USA}
\affiliation[d]{Enrico Fermi Institute,\\
The University of Chicago,  5640 South Ellis Avenue, Chicago, IL 60637, U.S.A.}
\affiliation[e]{Department of Physics,\\ University of Winnipeg, Winnipeg MB, R3B 2E9, Canada}

 \emailAdd{chowdh64@myumanitoba.ca}
 \emailAdd{ljenks3@jh.edu}
 \emailAdd{Rocky.Kolb@uchicago.edu}
 \emailAdd{e.mcdonough@uwinnipeg.ca}

\abstract{
We study cosmological gravitational particle production (CGPP) in Higgs inflation, wherein the inflaton is a scalar field with quartic self-coupling $\lambda$ and a nonminimal coupling to gravity $\xi$, and which may, but need not be, the Higgs boson of the Standard Model (SM).
We find an explosive particle production peaked on a characteristic comoving wavenumber $k\sim \xi^{2/3} a H$ with a peak occupation number that scales with $\xi$. This new peak in production can easily dominate over the conventional (minimally coupled inflation) CGPP even for modest values of $\xi$. The results apply for a wide range of $\xi$, e.g., as low as $\xi=10$, which can be realized for the Standard Model Higgs given suitable RG flow of the  quartic coupling. We discuss implications for late time relics such as dark matter.
}

\bottomimage{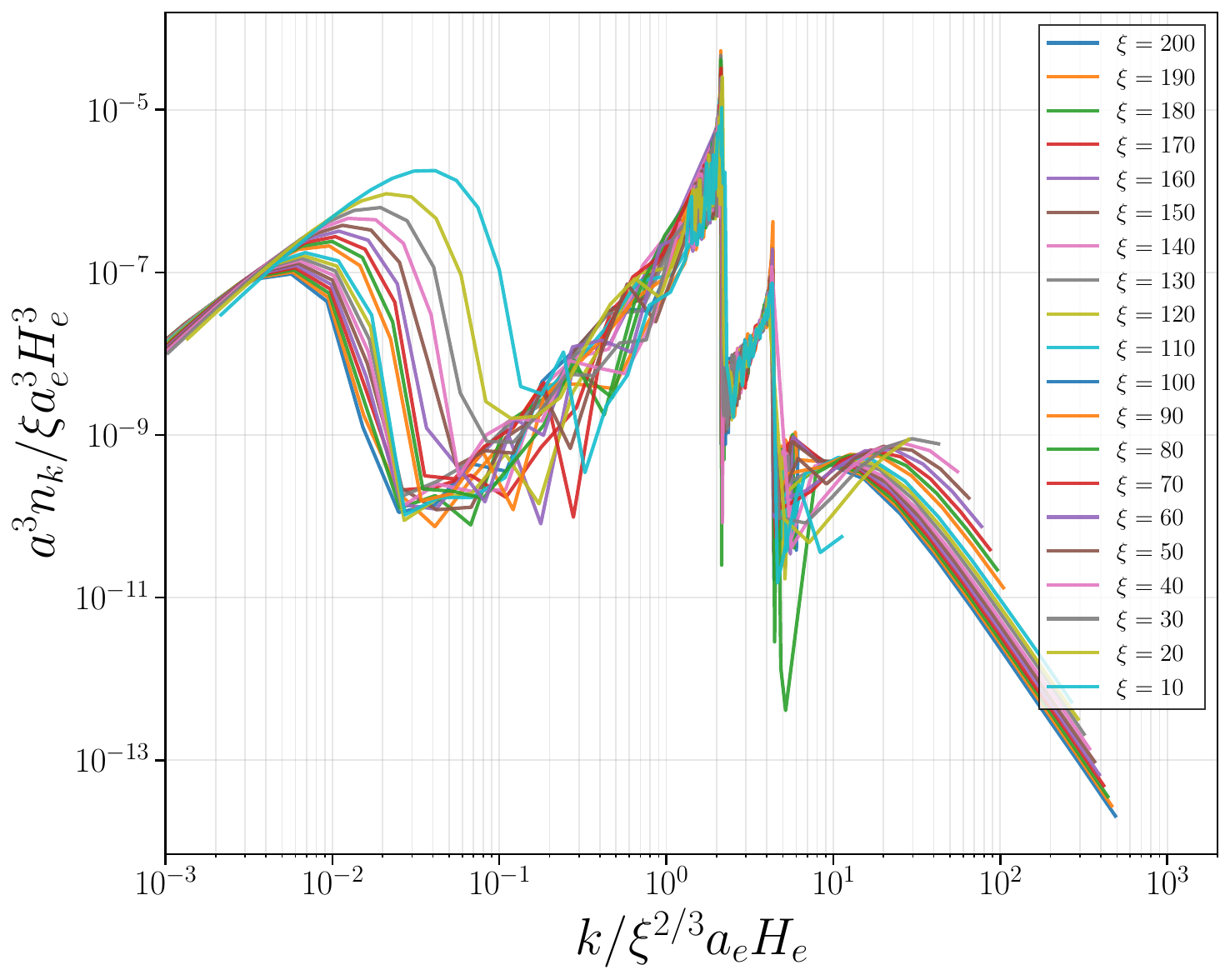}

\begin{document}
\maketitle
\flushbottom

\section{Introduction}
\label{sec:intro}

Cosmic inflation explains the observed homogeneity and flatness of the universe. It also provides a causal mechanism for structure formation, wherein the large-scale structure of the universe originates in quantum vacuum fluctuations. The key predictions of inflation are in excellent agreement with the observed universe \cite{Kallosh:2025ijd}, such as the prediction of a spectrum of primordial perturbations that is adiabatic, Gaussian, and nearly scale-invariant, in agreement with data from cosmic microwave background experiments \cite{Planck:2018jri}.

Many models that have been proposed to characterize the inflationary epoch (see e.g., \cite{Martin:2013tda} for a comprehensive list), are typically driven by one or more scalar fields as the inflaton. A natural possibility for the identity of the inflaton lies in the Higgs boson of the Standard Model \cite{Salopek:1988qh,Bezrukov:2007ep} (see Ref.~\cite{Rubio:2018ogq} for a review). This scenario avoids adding new degrees of freedom to explain inflation, and provides a natural connection between the Standard Model and inflation. In order to realize an extended phase of inflation without tuning the Higgs quartic self-coupling $\lambda$, the model relies on a nonminimal coupling of the Higgs to gravity, of the form $\xi \phi^2 R$ where $R$ is the spacetime Ricci scalar and $\xi$ is a (dimensionless) coupling constant. Couplings of this form have been intensely studied beginning a decade prior to the proposal of inflation, such as in Refs.~\cite{Chernikov:1968zm,Callan:1970ze,Bunch:1980br,Bunch:1980bs,Birrell:1982ix,Odintsov:1990mt,Buchbinder:1992rb,Parker:2009uva,Markkanen:2013nwa}. In these works it was shown that a nonminimal coupling is required for self-consistent renormalization of an interacting scalar field in curved spacetime. It is therefore well motivated to consider the cosmology of nonminimally coupled scalar fields, independent of whether the field is the Higgs boson of the Standard Model.

A fascinating and unavoidable outcome of cosmic inflation is the phenomenon of cosmological gravitational particle production (CGPP)  \cite{Parker:1969au, Parker:1971pt, Ford:2021syk,Chung:1998zb,Chung:1998ua} (see Ref.~\cite{Kolb:2023ydq} for a review). Originating in the work of Schr\"{o}dinger \cite{SCHRODINGER1939899}, CGPP results from the nonadiabatic expansion of spacetime. This provides a minimal mechanism to produce late-time relics, such as dark matter, from cosmic inflation. CGPP is ubiquitous in theories of inflation, including inflation models that derive  from string theory \cite{Ling:2025nlw} and those satisfying string swampland conjectures \cite{Kolb:2022eyn}, for a variety of spins and masses for the produced particles \cite{Chung:1998zb, Graham:2015rva, Ema:2015dka, Ema:2016hlw, Ema:2019yrd, Kolb:2020fwh,Ahmed:2020fhc, Alexander:2020gmv,Kolb:2021xfn,Kolb:2021nob,Kolb:2023dzp, Capanelli:2023uwv,Kaneta:2023uwi,Maleknejad:2022gyf,Capanelli:2024rlk,Capanelli:2024pzd,Jenks:2024fiu}.

In this work we study the CGPP of scalar fields in the context of Higgs inflation, where the inflaton is a nonminimally coupled scalar field with a quartic self-interaction, and which may be, but need not be, identified with the SM Higgs boson. CGPP has been studied in this context previously in Refs.~\cite{Ema:2016dny,Babichev:2020yeo} and related work Ref.~\cite{Karam:2020rpa}, but the spectrum of produced particles in Higgs inflation has never been directly computed or presented. We find qualitatively new features in CGPP arising from previously underappreciated and unknown features in the post-inflationary background evolution of Higgs inflation and nonminimally coupled inflation generally.   This establishes Higgs inflation as a {\it particle factory}, capable of producing new particles in abundance, which can be put to a variety of uses, such as dark matter.

Beginning with the background evolution, we find a new universal scaling of the slow-roll parameter $\varepsilon=-\dot{H}/H^2$ with the nonminimal coupling $\xi$, namely that $\varepsilon = 3 +6 \xi$ whenever the inflaton $\phi$ passes through $0$. This simple analytic relation explains `spikes' in field velocity and sharp features in the Hubble parameter. We also find that, consistent with past work in the context of preheating \cite{DeCross:2015uza,DeCross:2016fdz,DeCross:2016cbs,Nguyen:2019kbm,vandeVis:2020qcp,Sfakianakis:2018lzf}\footnote{For a review of preheating, see e.g.~Ref.~\cite{Amin:2014eta}.}, inflation exits to a phase with equation of state $w=0$, wherein the frequency of inflaton oscillations is roughly constant,  and later transitions to $w=1/3$, as expected for a field oscillating in a quartic potential, at which point the frequency of inflaton oscillations begins to redshift. The duration of the phase with $w=0$ depends on $\xi$, with larger $\xi$ leading to a prolonged phase of $w \approx 0$ and hence a delayed onset of redshifting of the frequency of inflaton oscillations. This is imprinted on the late-time frequency of inflaton oscillations, $\omega_\phi$, which converges to a universal scaling regime like $\omega_\phi \propto \xi^{2/3}$.

These new features in the background evolution lead to new features in CGPP: we find a sharp peak in the spectrum of particle production for comoving wavenumber $k_{\rm peak}\sim\xi^{2/3} a_e H_e$, and secondary peak at $k=2k_{\rm peak}$, with an occupation number that scales linearly with nonminimal coupling of the inflaton,  $n_{k_{\rm peak}}\propto \xi$.  This new feature is exhibited for both conformally coupled and minimally coupled spectator scalars, across a broad range of Higgs nonminimal coupling $\xi$ and spectator mass $m_{\chi}$. Importantly, from the time-evolution of occupation number, it is readily apparent that the asymptotic limit of $n_k$, which defines the late time particle number density, is the cumulative result of many successive oscillations of the background, and cannot be attributed to the first `spike' in field velocity or the Ricci scalar.

The consequent integrated particle number density can be significantly larger than conventional minimally coupled inflation models, illustrating the power of Higgs inflation to serve as a {\it particle factory}. We apply this particle factory to the problem of dark matter, and identify two related but distinct paths to obtain the observed dark-matter relic density: we consider the possibility that the produced particles are stable and constitute the observed dark matter, or alternatively, that the produced particles decay, and the decay products are the dark matter. In both cases we are able to match the observed abundance of dark matter.

The paper is structured as follows; in Sec.~\ref{sec:Higgs} we provide an overview of Higgs inflation. In Sec.~\ref{sec:ABCs}, we briefly review gravitational particle production. Section~\ref{sec:HiggsGPP} presents our results for the particle production of both conformal and minimal couplings in Higgs inflation. In Sec.~\ref{sec:latetimerelics} we analyze the application of CGPP to generating the observed DM density. Finally, in Sec.~\ref{sec:Discussion}, we provide our discussion and conclusions.

\section{Higgs Inflation}
\label{sec:Higgs}

For cosmological purposes one may model the SM Higgs as a real scalar field, with the action given by
\begin{equation}
 S = \int d^4 x \sqrt{ - g} \bigg[ -\frac{1}{2}M_\mathrm{Pl}^2 R - \frac{1}{2} \left( \partial \phi \right)^2    - \frac{ \lambda}{4} \left( \phi^2 - v^2 \right)^2 - \frac{1}{2}\xi \phi^2 R \bigg].
\label{eq:actionHiggsInflation}
\end{equation}
The scalar field $\phi$ need not be identified with a component of the Higgs boson of the Standard Model, but it is nonetheless an exciting and interesting prospect, and thus we briefly consider this possibility further.

The identification of $\phi$ with SM Higgs allows the parameters $v$ and $\lambda$ to be determined experimentally by the measured values of the Higgs VEV and quartic self coupling, $v=246$ GeV and $\lambda = m_h^2/(2 v^2)=0.1$. On the other hand, matching Higgs inflation to CMB observables fixes the parameter combination $\xi \simeq 50000 \sqrt{\lambda}$ \cite{Bezrukov:2007ep}, and therefore $\lambda=0.1$ implies $\xi \sim 10^4$. Such a large value of $\xi$ presents a problem, since the nonminimal coupling rescales the UV cutoff of gravity from the Planck scale to $M_\mathrm{Pl}/\xi \sim 10^{14}$ GeV \cite{Hertzberg:2010dc}, which is near the Hubble parameter during Higgs inflation, $H_{\rm inf} \sim 10^{13}$ GeV, suggesting a breakdown of perturbative unitarity for quantum fluctuations in Higgs inflation. Since Ref.~\cite{Hertzberg:2010dc}, the issue of perturbative unitarity in Higgs inflation has been intensely studied \cite{Burgess:2010zq,Barbon:2009ya,Giudice:2010ka,Lerner:2011it,Barbon:2015fla,Steingasser:2025txd}.

However the argument presented above neglects the renormalization group flow of $\lambda$. The properties of the Higgs are well measured at the TeV scale, whereas the characteristic energy scale of Higgs inflation is $\mathcal{O} (10^{13}$ GeV). Even neglecting new particles, the Higgs quartic coupling flows to small values at high energies and can even flow to negative values.  New particles beyond the Standard Model can alter the RG flow (see e.g.~\cite{Steingasser:2023ugv}), leading to a considerable theory uncertainty on the value of $\lambda$ at high energies. Since $\lambda$ and $\xi$ are connected by the CMB constraint $\xi \simeq 50000 \sqrt{\lambda}$, the freedom to set $\lambda$ amounts to a freedom to set $\xi$, and therefore ample opportunity to avoid issues with perturbative unitarity. For example, $\xi=10$ can provide a suitable realization of SM Higgs inflation\footnote{
We note that Higgs inflation  with  a modest value of $\xi$ and hence $\lambda \ll 1$ can be sensitive to radiative corrections to the inflationary potential, and the corrected model is often referred to as critical Higgs inflation \cite{Bezrukov:2014bra,Hamada:2014iga,Hamada:2014wna}. In this work we restrict our attention to the tree-level potential, and note critical Higgs inflation as an interesting direction for future work. }. Guided by this, in what follows we will consider Higgs-like inflation for a real scalar $\phi$ and a nonminimal coupling $\xi$, with $\xi$ a free parameter, fixed only by the CMB constraint on $\xi$ and $\lambda$.

The action, Eq.~\eqref{eq:actionHiggsInflation}, is defined in the Jordan frame, which makes the nonminimal coupling manifest. A Weyl transformation to the Einstein frame can remove the coupling term, but makes characterizing the effect of the nonminimal coupling on particle production more complicated. Therefore, in what follows we will remain in the Jordan frame\footnote{
Other works which work directly in the Jordan frame include Refs.~\cite{Kaiser:2010yu,Figueroa:2021iwm}.}. From Eq.~\eqref{eq:actionHiggsInflation}, and assuming a Friedman-Lema\^{i}tre-Robertson-Walker background\footnote{We adopt a ``mostly-minus'' metric signature but note that the equations of motion are invariant under change of metric signature. See App.~\ref{app:sign_conventions} for a detailed derivation.} 
\be 
{\rm d}s^2 = {\rm d}t^2 - a(t)^2 ({\rm d}x^2 + {\rm d}y^2 + {\rm d}z^2)
\ee 
the equations of motion are given by a modified Klein-Gordon equation for $\phi$, 
\begin{equation}
\label{eq:phiEOM}
    \ddot{\phi}+ 3 H \dot{\phi} - 6  \xi \phi \left( H^2 + \dot{H} \right) + \frac{dV}{d\phi}=0
\end{equation}
where $H=\dot{a}/a$, and a modified Friedmann equation, 
\begin{equation}
\label{eq:HEOM}
     3   M_\mathrm{Pl}^2 H^2 + \xi \phi^2 \left( 3 H^2  + 6 H  \frac{d \log \phi}{dt}\right) = \frac{1}{2}\dot{\phi^2} + V(\phi) .
\end{equation}
From this one may infer $\dot{H}$, which satisfies 
\begin{equation}
    (M_\mathrm{Pl}^2+\xi\phi^2)(2\dot{H}+3H^2)+2\xi(\phi\ddot{\phi}+\dot{\phi}^2+2H\phi\dot{\phi})=-\frac{1}{2}\dot{\phi}^2+V(\phi) ,\label{eq: Hdot}
\end{equation}
which can be derived directly from the action by using the ADM formalism, as demonstrated explicitly in App.~\ref{app:sign_conventions}.

Inflation occurs in the regime where the nonminimal coupling term is large compared to the Planck scale, ${\xi} \phi^2 \gg M_{\rm pl}^2$. A detailed discussion of the CMB observables in Higgs inflation can be found in, e.g., \cite{Bezrukov:2007ep,Rubio:2018ogq}. For $\xi\gg1$, the spectral index $n_s$ and the tensor-to-scalar ratio $r$ are independent of $\lambda$ and $\xi$, while the amplitude of the primordial power spectrum of curvature perturbations, $A_s$, does depend on $\lambda$ and $\xi$, which imposes the relation
\begin{equation}
    \xi \simeq 5\times10^4 \sqrt{\lambda}. 
\end{equation}
This constraint fixes $H \sim 10^{13}$ GeV during inflation. Inflation ends when  $\sqrt{\xi}\phi \simeq  M_\mathrm{Pl} $, at which point  $\phi$ undergoes damped oscillations similar to conventional minimally coupled inflation models.  

To illustrate these dynamics, we numerically solve the system of equations Eqs.~\eqref{eq:phiEOM} and \eqref{eq:HEOM} for a set of fiducial $\xi$ and with $\lambda$ set to satisfy the $A_s$ constraint (we note that $\lambda$ scales out of all quantities once they are expressed in units of Hubble at the end of inflation, $H_e$). We impose an initial condition that $\sqrt{\xi}\phi_i \gg M_\mathrm{Pl}$ to realize a long-lived phase of slow-roll inflation. We set the Higgs VEV parameter $v=0$ now and hereafter, since the observed Higgs VEV, $v\approx 10^{-16} M_{\rm pl}$ is too small to play any role in the cosmological dynamics during or after inflation.

\begin{figure}[htb!]
\centering 
    \includegraphics[width=1\textwidth]{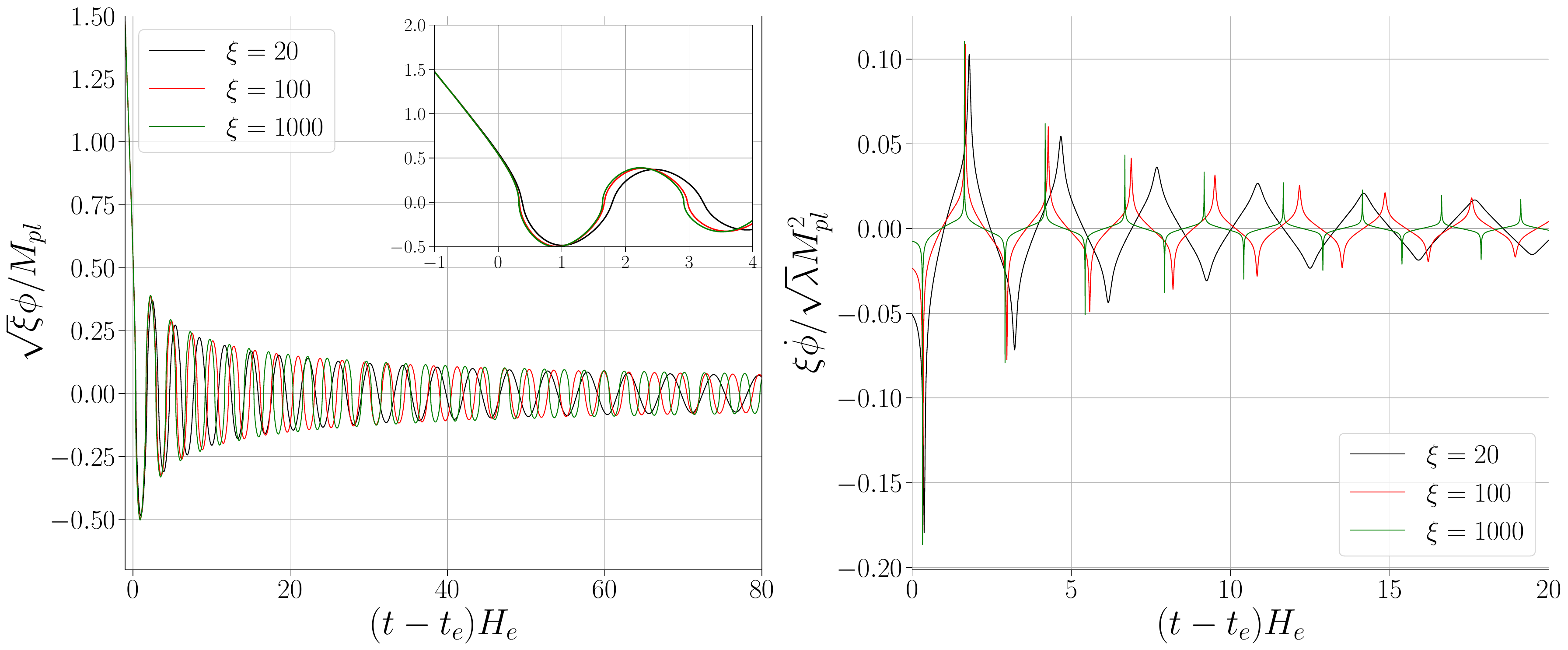}
    \caption{
    Evolution of the inflaton $\phi$ for three fiducial values of $\xi$ with respect to time in the Jordan frame is shown in the left panel. The inset shows the cohesion of frequencies for large values of $\xi$ at early times. At late times, the inflaton begins to oscillate at different rates regardless of the value of $\xi$ and the frequencies begin to slow down with time. In all cases $\lambda$ is rescaled to match to the CMB $A_s$ constraint. The right panel shows the evolution in $\dot{\phi}$, featuring sharp spikes.}
    \label{fig:phi_evol}
\end{figure}

\Cref{fig:phi_evol} shows the post-inflationary evolution of the field $\phi$ and the field velocity $\dot{\phi}$, in the left and right panel respectively. The field and field velocity have been rescaled by $\sqrt{\xi}$ and $\xi$ respectively.  From the left panel, one may appreciate that the scaling symmetry of $\phi$ with $\sqrt{\xi}$ continues at early times in the post-inflation oscillatory phase, and for $\xi \gg1$ this endures for longer. From the right panel, one may appreciate that the field velocity exhibits sharp spikes, up to ${\rm max}(\dot{\phi} )\sim 0.1 (\sqrt{\lambda}/\xi) M_\mathrm{Pl}^2$. After imposing the CMB $A_s$ constraint, these spikes have a universal amplitude, ${\rm max}(\dot{\phi})\sim 10^{-6} M_\mathrm{Pl}^2$. The spikes occur near the zero-crossings of $\phi$, and have been discussed in, e.g.,~\cite{Ema:2016dny,Babichev:2020yeo}.

\begin{figure}[htb!]
\centering 
    \includegraphics[scale=0.5]{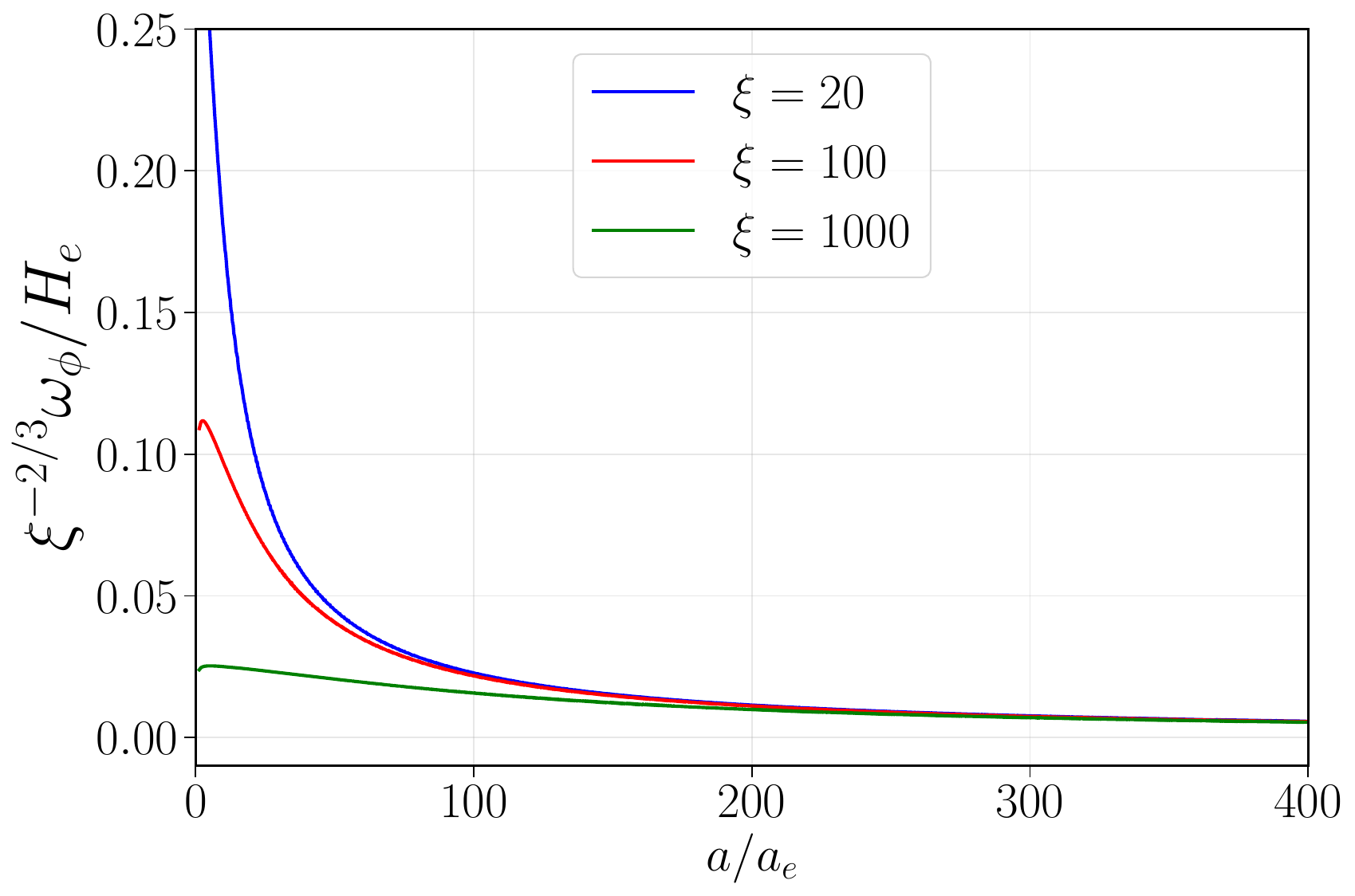}
    \caption{Scaling of the oscillation frequencies of the inflation at late times. We define the cycle-averaged frequency as the inverse period of oscillations. At late times the frequency approaches a universal scaling solution which scales as $\xi^{2/3}$. }
    \label{fig:oscill_scaling}
\end{figure}

The scaling symmetry in $\xi$ is ostensibly lost at late times after inflation, as can be appreciated from Fig.~\ref{fig:phi_evol}. However a new scaling emerges, this time in the {\it frequency} of oscillations. We define the time-averaged frequency $\omega_\phi = \pi/\Delta t$ using the time interval between inflaton zero-crossings $\Delta t$. In Fig.~\ref{fig:oscill_scaling} we plot the evolution of $\omega_\phi$,  scaled by $\xi^{2/3}$. From this one may appreciate that the oscillations rapidly approach a scaling regime wherein $\omega_\phi \propto \xi^{2/3}$.

\begin{figure}[htb!]
\centering 
    \includegraphics[scale=0.5]{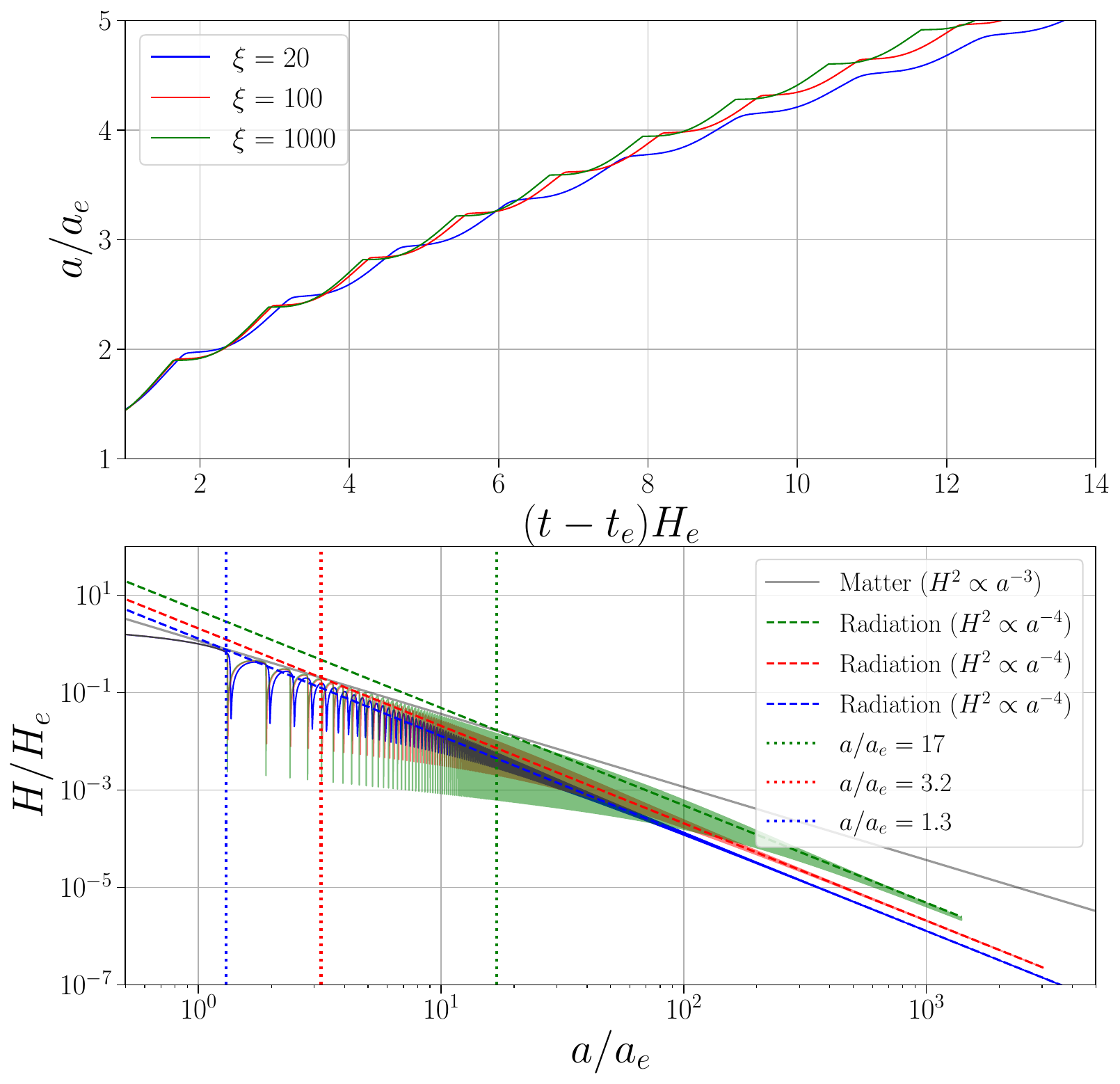}
    \caption{ Evolution of the comoving scale factor $a/a_e$ with time shows cuspy oscillations. In the bottom panel, the evolution of the Hubble scale is shown with respect to $a/a_e$. Dashed lines show the evolution of a radiation-dominated universe. The time at which $H$ evolves like radiation is distinguished with vertical dotted lines, for three different values of $\xi$.}
    \label{fig:H_a_plot}
\end{figure}

The evolution of $a$ and $H$ are shown in Fig.~\ref{fig:H_a_plot}.  The nonminimal coupling is encoded into the background in two ways:
\begin{enumerate}
    \item Cuspy oscillations in the scale factor $a(t)$, shown in the top panel of Fig.~\ref{fig:H_a_plot}. These cuspy features are translated into sharp features in the Hubble parameter $H/H_e$ shown in the bottom panel. 
    \item The evolution interpolates from effectively matter dominated ($H\sim a^{-3/2}$) at early times to radiation dominated ($H \sim a^{-2}$) at late times. These two distinct regimes correspond to periods in time where the nonminimal coupling dominates the evolution and when the quartic potential dominates the evolution, respectively. This is consistent with previous studies of nonminimally coupled inflation models, c.f., Fig.~9 of \cite{DeCross:2015uza}.
\end{enumerate}
The second of these in part explains the $\xi^{2/3}$ scaling of the frequency shown Fig.~\ref{fig:oscill_scaling}. At early times the frequency is nearly constant and is ${\cal O}(H_e)$ independent of $\xi$, whereas at late times, when the model effectively evolves as a minimally coupled scalar in a quartic potential, the oscillation frequency redshifts as in a $\xi$-independent way, namely as $\omega_\phi \propto a^{-2}$. The transition between these two regimes (constant vs. redshifting frequency) is controlled by $\xi$, with the transition occurring later for larger $\xi$, leading to an overall scaling with $\xi$ as $\omega_\phi \propto \xi^{2/3}$. 

Finally, we return to the `spikes' in the field velocity. This can be understood analytically by examining the slow-roll parameter $\varepsilon \equiv - \dot{H}/H^2$ at the zero-crossings of the inflaton. In conventional inflation models, $\varepsilon(\phi=0)=3$. In Higgs inflation, as one may easily find by setting $\phi=\ddot{\phi}=0$ in the background equations of motion, the slow-roll parameter at zero-crossings of $\phi$ is given by:
\begin{equation}
    \varepsilon(\phi=0) = 3 + 6 \xi .
\end{equation}
This indicates that $\xi \gg 1$ will experience a dramatic increase in $\dot{H}$ as $\phi$ passes through the origin. This leads to features in $\dot{\phi}$ and $H$ that have been noted previously \cite{Ema:2016dny,Babichev:2020yeo} but never derived exactly.

\begin{figure}[htb!]
\centering 
    \includegraphics[width=0.49\textwidth]{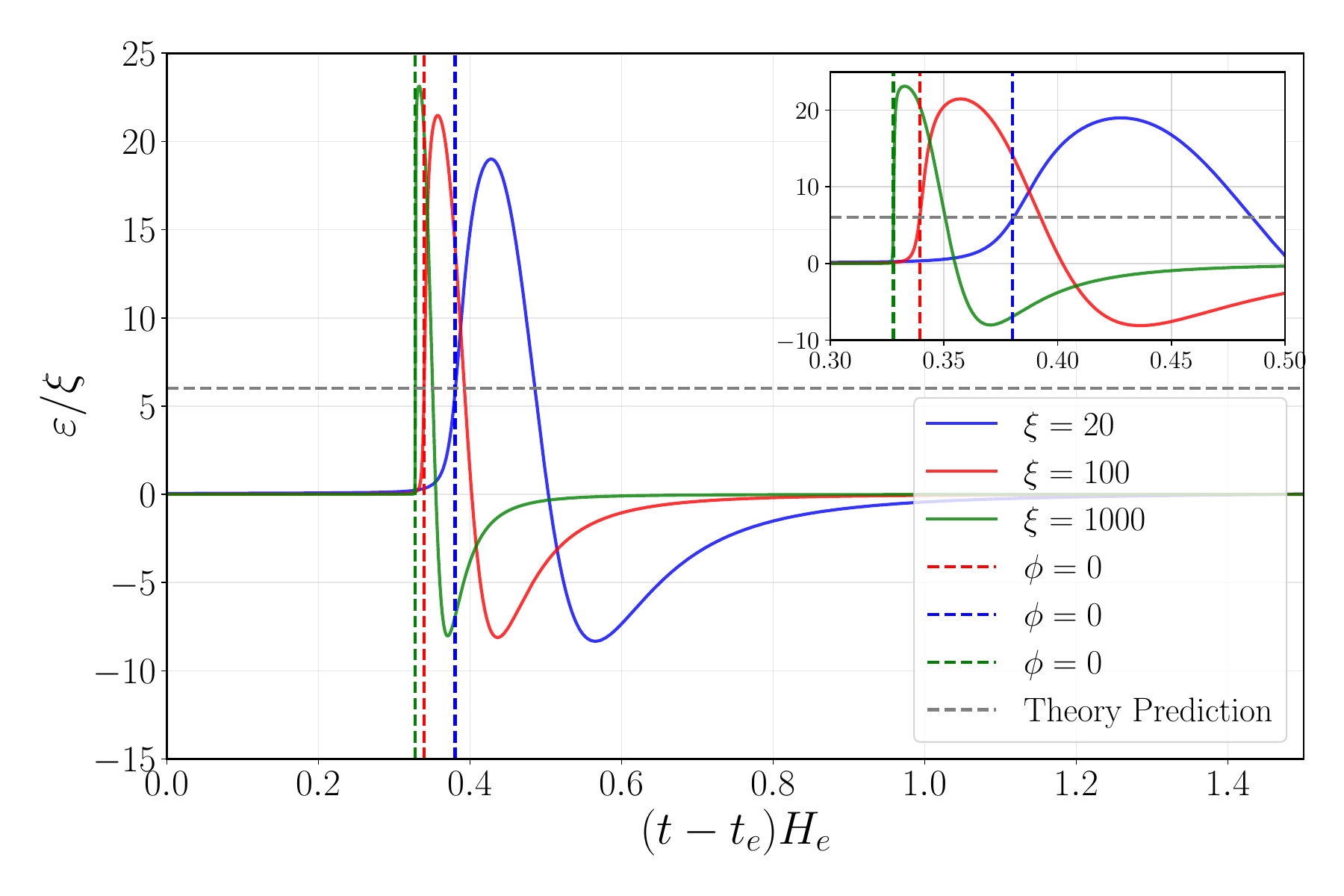}
    \includegraphics[width=0.49\textwidth]{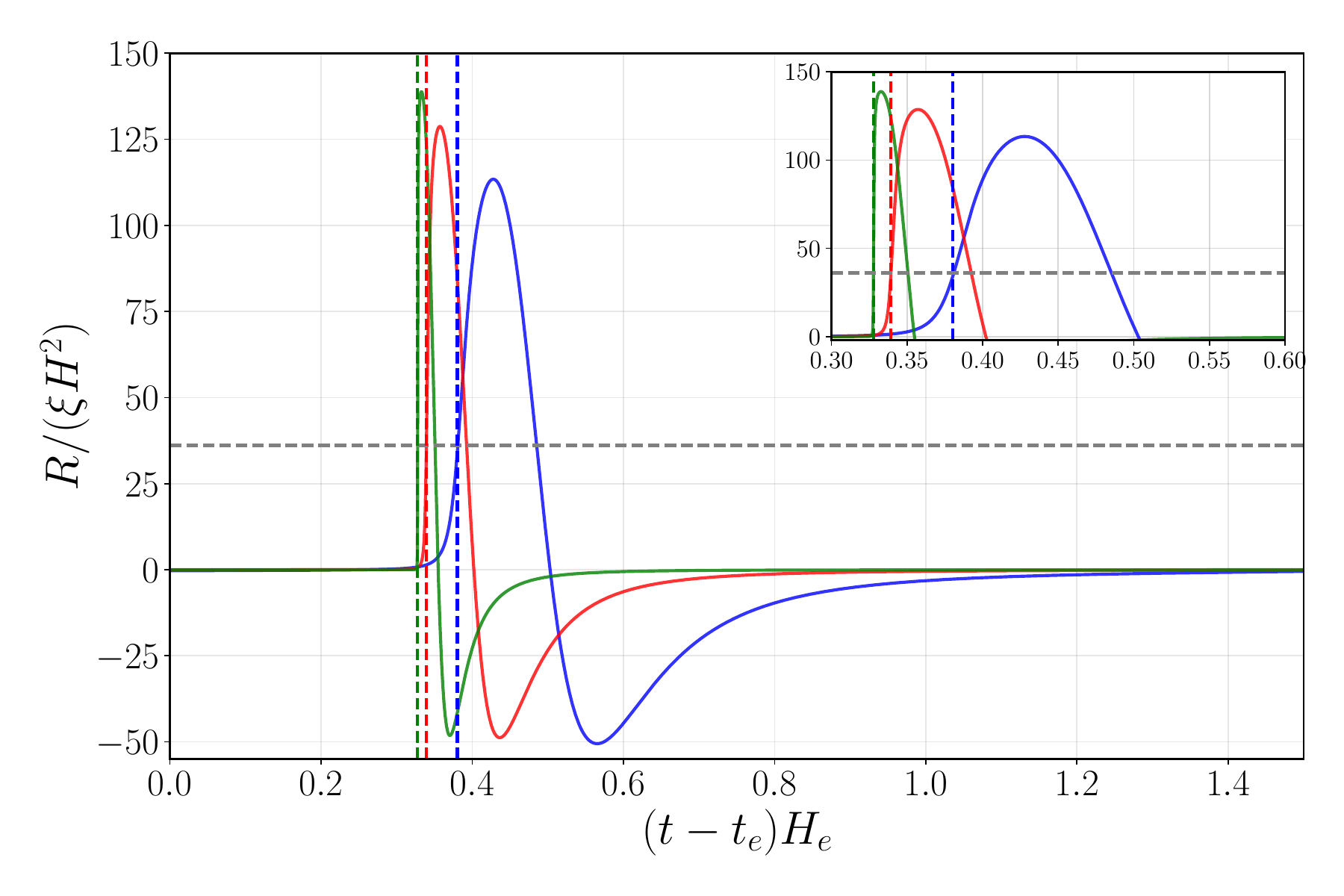}
    \caption{ 
    Evolution of the slow-roll parameter $\varepsilon$ (left panel), scaled by $\xi$, and of the Ricci scalar $R$ (right panel), scaled to $\xi H^2$. At the end of inflation $\varepsilon/\xi=1/\xi$, i.e., $ \varepsilon=1$, which rapidly grows to $\varepsilon/\xi \sim 6$, i.e., $\varepsilon\sim 6\xi$ when $\phi$ passes through zero. This differs from conventional inflation models, where $\varepsilon(\phi=0)=3$. Similarly, the Ricci scalar is $R \sim 36 \xi H^2$ at $\phi=0$, in comparison with conventional inflation models where $R(\phi=0)=6 H^2$. We note that, relative to $H_e^2$, the peak value of $R$ is ${\cal O}(1)\xi$, i.e., max$(R)={\cal O}(1)\xi H_e^2$ (see Fig.~\ref{fig:minimal-ricci}). } 
    \label{fig:epsilon}
\end{figure}

\begin{figure}[ht!]
\centering 
    \includegraphics[scale=0.3]{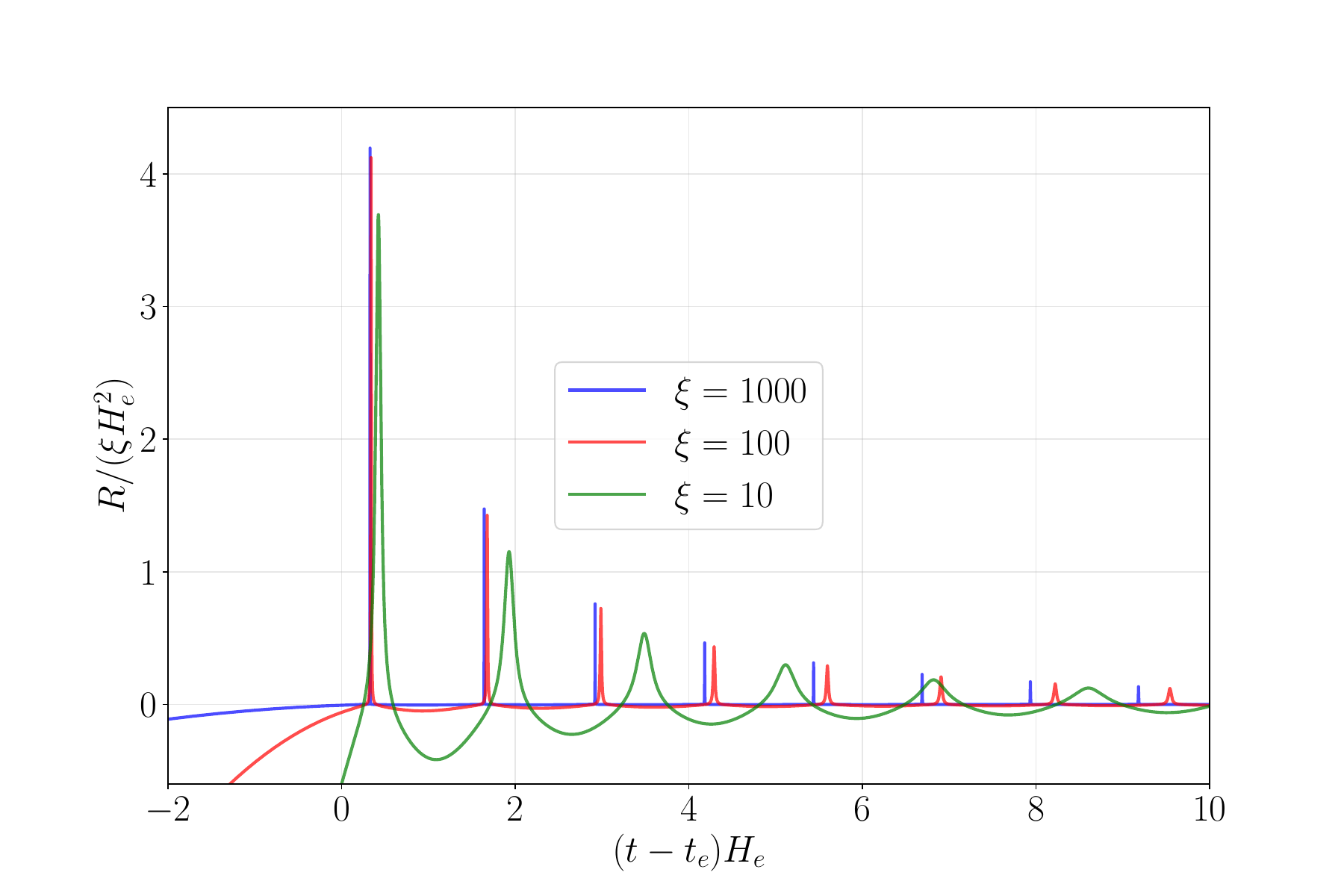}
    \caption{Time evolution of the Ricci scalar in Higgs inflation, scaled to $\xi H_e ^2$.}
    \label{fig:minimal-ricci}
\end{figure}

The spikes in $\varepsilon$ are encoded into the Ricci scalar, which can be expressed as $R= - 6 H^2 (2-\varepsilon)$, such that $R$ at zero-crossings of the inflaton is given by
\begin{equation}
    R(\phi=0) = 6H^2(1+ 6\xi).
\end{equation}
In Fig.~\ref{fig:epsilon} we confirm our predictions for $\varepsilon$ and $R$. A red dashed line shows the predicted value of $\varepsilon$ when $\phi=0$ (shown in blue, green and orange dashed lines for $\xi=20,100,1000$, respectively).  From the right panel, which shows the evolution of $R(t)/H(t)^2$ scaled by $\xi$, one may appreciate that the theory prediction is again confirmed. We note that when expressed in units of $H_e$, the spikes in $R$ reach a maximum amplitude ${\rm max}(R)= {\cal O}(1) \xi H_e ^2$ for the first spike and redshifting thereafter. This is shown in Fig.~\ref{fig:minimal-ricci}, and will be important for interpreting the production of the minimally coupled spectator.

\section{ABCs of CGPP: Review of Gravitational Particle Production}
\label{sec:ABCs}

Having studied the dynamics of our particular Higgs inflation model, let us now turn to a brief overview of cosmological gravitational particle production. For further details, see \cite{Kolb:2023ydq} and references therein. We will consider the gravitational production of an additional scalar (not the inflaton) that acts as a spectator field during inflation.

We consider a spectator field not directly coupled to the inflaton, and constituting a negligible fraction of the energy density of the universe during inflation. We apply this definition in the Jordan frame,\footnote{The equivalence of CGPP in the Jordan and Einstein frames has been shown in Ref.~\cite{Ema:2016dny}.} where the action of the inflation model and gravity is given by Eq.~\eqref{eq:actionHiggsInflation}. The action for the spectator scalar field is given by:
\begin{equation}
    S[\varphi(\eta,\textbf{x})]=\int d\eta ~ d^3\textbf{x}\left[\frac{1}{2}a^2(\partial_{\eta}\varphi)^2-\frac{1}{2}a^2(\nabla\varphi)^2-\frac{1}{2}a^4m^2\varphi^2+\frac{1}{2}a^4\bar{\xi} R\varphi^2\right],
\end{equation}
where $\varphi$ is the spectator field, $\eta$ is conformal time, and $\bar{\xi}$ is the nonminimal coupling between gravity and the spectator field. Note that $\bar{\xi}$ is distinct from the inflaton nonminimal coupling $\xi$ used in the previous section. In the context of particle production, it is common to consider spectator fields that are conformally coupled ($\bar{\xi} =1/6$) and minimally coupled ($\xi = 0$) to gravity, and in the subsequent section we will analyse both cases in detail.

In order to normalize the kinetic term, we perform a field redefinition $\chi(\eta,\textbf{x})=a(\eta)\varphi(\eta,\textbf{x})$ resulting in 
\begin{equation}
    S[\chi(\eta,\textbf{x})]=\int d\eta~d^3\textbf{x}\left[\frac{1}{2}\left(\partial_{\eta}\chi\right)^2-\frac{1}{2}\left(\nabla\chi\right)^2-\frac{1}{2}a^2m_{\text{\rm eff}}^2\chi^2\right]
\end{equation}
where we have dropped a total derivative term that vanishes when $\eta\rightarrow{\pm\infty}$. The effective mass is
\be 
 m_{\rm eff}^2\equiv m^2+\frac{1}{6}(1-6\bar{\xi})R.
\ee
The spectator field, $\chi$ can be decomposed into Fourier modes as
\begin{equation}
    \hat{\chi}(\eta,\textbf{x})=\int\frac{d^3\textbf{k}}{(2\pi)^3}\left[\hat{a}_{\textbf{k}}\chi_{\textbf{k}}(\eta)e^{i\textbf{k}\cdot \textbf{x}}+\hat{a}_{\textbf{k}}^{\dagger}\chi_{\textbf{k}}^*(\eta)e^{-i\textbf{k}\cdot \textbf{x}}\right]
\end{equation}
where $k = |\textbf{k}|$ is the wavenumber and $\hat{a}_k^{\dagger}$ and $\hat{a}_k$ are the creation and annihilation operators respectively. This field is a solution to the equation of motion given by
\begin{equation}
    \partial_{\eta}^2\chi_k(\eta)+\omega_k^2(\eta)\chi_k(\eta)=0
    \label{eq:mode_eq}
\end{equation}
where the $\omega_k^2$ is given by  
\begin{equation}
    \omega_k^2= k^2+a^2(\eta)m^2+\frac{1}{6}(1-6\bar{\xi})a^2(\eta)R(\eta). 
\end{equation}
When solving for the mode functions we assume Bunch-Davies initial conditions given by
\begin{equation}
    \chi_{k_0}(\eta)\equiv\frac{1}{\sqrt{2k}}e^{-ik\eta}~, ~~~ \partial_{\eta}\chi_{k_0}(\eta)=-i\sqrt{\frac{k}{2}}e^{-ik\eta}~,
\end{equation}
imposed in the limit $ \eta\rightarrow-\infty$.

From the mode functions one may construct the Bogoliubov coefficient,
\begin{equation}
    |\beta_k|^2=\frac{1}{\omega_k}\left(\frac{1}{2}|\partial_{\eta}\chi_k|^2+\frac{1}{2}\omega_k^2|\chi_k|^2\right)-\frac{1}{2}~.
\end{equation}
The comoving particle number density is then given by,
\begin{equation}
     a^3 n = \int a^3 n_k \, d\log k
     \label{eq:na3}
\end{equation}
where $n_k$, defined by
\begin{equation}
\label{eq:nk}
    a^3 n_k= \lim_{\eta \rightarrow \infty} \frac{k^3}{2\pi^2}|\beta_k|^2,
\end{equation}
is the particle number per logarithmic decade in $k$.

\section{Particle Production in Higgs Inflation}
\label{sec:HiggsGPP}

\begin{figure}[ht!]
\centering 
    \includegraphics[scale=0.5]{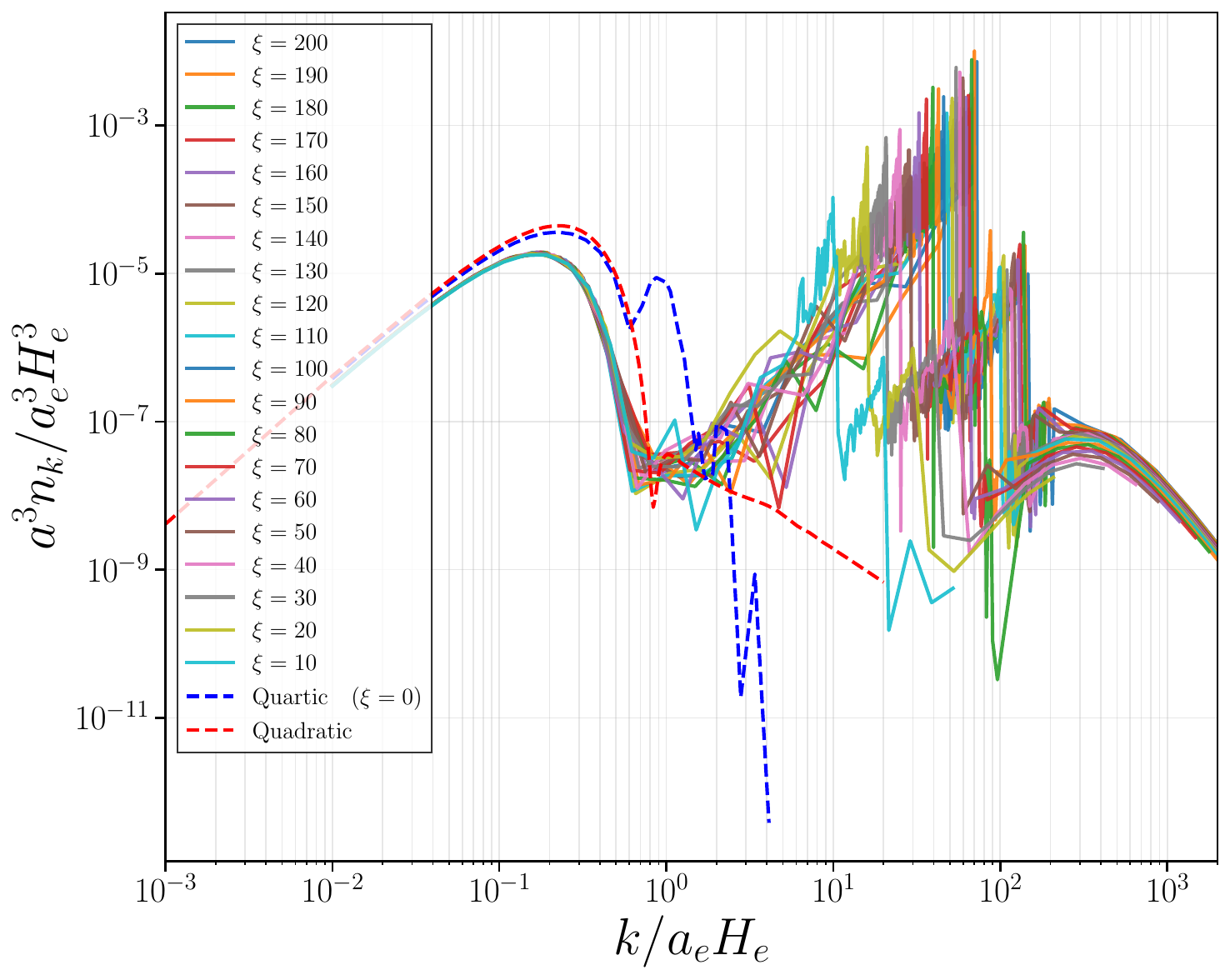}
    \caption{
    Particle production in Higgs inflation for a conformally coupled spectator field with mass $m_{\chi}/H_e=0.1$, for various values of the Higgs nonminimal coupling $\xi$ (solid curves). This is compared with a quartic potential where $\xi=0$ in a blue dashed line. A peak in $n_k$ is also observed in the quartic case, however significantly less sharp and pronounced than those of Higgs inflation. In a red dashed line, the spectrum for the quadratic potential is shown. For all three cases (Higgs, quartic and quadratic), the spectrum follows the same growth in $n_k$ for low $k$.  
    } 
    \label{fig:spectrum_no_rescaling}
\end{figure}

\begin{figure}[ht!]
\centering 
    \includegraphics[scale=0.5]{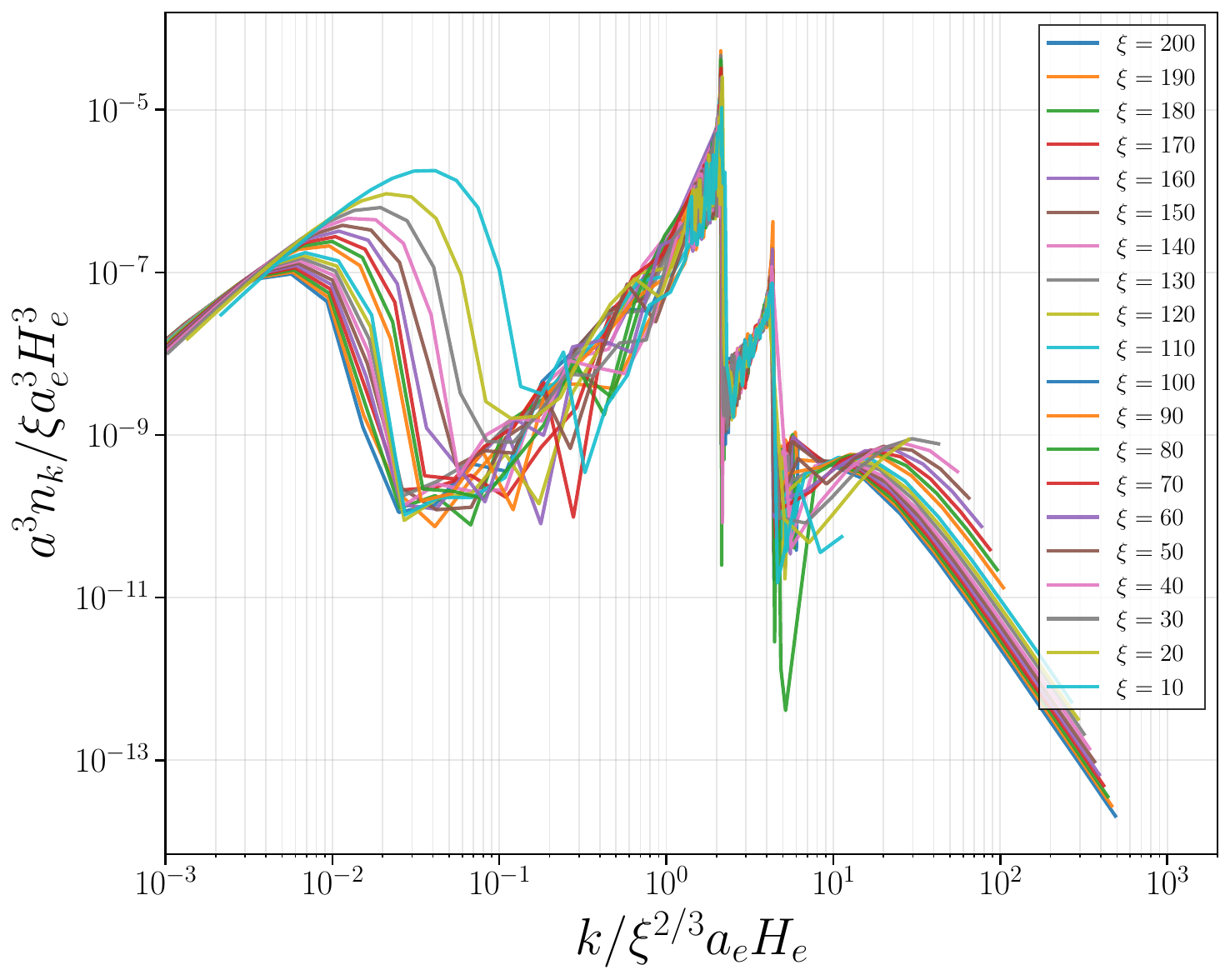}
    \caption{Particle production for Higgs inflation for $m_{\chi}/H_e=0.1$ and $\xi$ ranging from $10-200$, with $k$ scaled by $\xi^{2/3}$ and $n_k$ scaled by $\xi$. This shows a near-perfect alignment in $k_{peak}$ and $2k_{peak}$ for all values of $\xi$. 
    } 
    \label{fig:spectrum_rescaled}
\end{figure}

Let us now study the dynamics of CGPP in Higgs inflation. We numerically solve the mode equations, Eq.~\eqref{eq:mode_eq} to find $|\beta_k|^2$ and the resulting particle number, $n_k$, for various parameters of the model\footnote{We work to linear order in perturbation theory, but note that a full nonlinear analysis, for example, simulations on a lattice akin to that performed in the context of preheating, such as in Refs.~ \cite{Nguyen:2019kbm,vandeVis:2020qcp}, would be an important next step.}$^,$\footnote{While we pay particular attention to $\xi$, the nonminimal coupling of the {\it inflaton}, we note a significant body of work focusing on the nonminimal coupling of the field undergoing particle production, denoted by $\bar{\xi}$ in our work, such as Refs.~\cite{Opferkuch:2019zbd,Bettoni:2021zhq,Laverda:2023uqv,Figueroa:2024asq,deHaro:2024xgd,Fairbairn:2018bsw,Clery:2022wib,Verner:2024agh} and the review \cite{Kolb:2023ydq}.}. Our key science results are the spectra shown in Figs.~\ref{fig:spectrum_no_rescaling},~\ref{fig:spectrum_rescaled},~\ref{fig:spectrum_mass}, and~\ref{fig:spectrum_minimum}. 

Fig.~\ref{fig:spectrum_no_rescaling} shows the spectrum of produced particles for a conformally coupled scalar ($\bar{\xi}=1/6$) with mass $m_{\chi}=0.1 H_e$ where $H_e$ is the Hubble scale at the end of inflation, in Higgs inflation with varying values of $\xi$. Quadratic inflation and quartic inflation are also shown in dashed lines for comparison. At low $k$ ($k< a_e H_e$), corresponding to modes that exited the horizon already during inflation, the spectrum exhibits the characteristic peak of particle production seen in minimally ($\xi=0$) coupled inflation models such as quadratic inflation and quartic inflation. At low $k$, there is an increase in the number density which scales as $a^3n_k/a_e^2H_e^2 \sim k^2$ for all three models. The spectrum in quadratic inflation has a characteristic exponential drop followed by a $k^{-3/2}$ fall off at large $k$ \cite{Kolb:2023ydq} (red dashed line). Similarly, the quartic inflation (red dashed line) spectrum has some secondary oscillations, then has a sharp exponential fall off after the initial peak. 

The spectra of Higgs and minimally coupled inflation differ dramatically at high-$k$ ($k> a_e H_e$). The Higgs inflation spectra exhibit a new peak in the particle production at $k_{\rm peak} \simeq 2\xi^{2/3} a_e H_e$  and a secondary peak at $k=2k_{\rm peak}$, in addition to the usual characteristic peak in particle production that one sees for CGPP in simpler models such as quadratic inflation. The amplitude of these new features scale linearly with $\xi$. Beyond $k\sim 2k_{\rm peak}$, the spectrum decays as $k^{-3/2}$ as in quadratic inflation.

To illustrate the scaling behavior of the spectra, in Fig.~\ref{fig:spectrum_rescaled} we show a spectrum with $k$ rescaled by $\xi^{2/3}$ and $n_k$ rescaled by $\xi$. From this one can see the near-perfect alignment of the sharp features at $k=k_{\rm peak}, 2k_{\rm peak}$, as well as the constancy of the scaling of the peak amplitude. The emergence these new features and their striking universality are the main hallmark of gravitational particle production in Higgs inflation.

To understand these results, it is helpful to return to the background evolution. As we have discussed, the coupling of the inflaton to gravity induces cuspy oscillations in the scale factor, $a$, and sharp oscillations in the Hubble parameter, $H$. For a conformally coupled spectator field, the oscillations in $a$ induce the same cuspy oscillations in $\omega_k^2$, which controls the particle production. For minimally coupled scalars, there are sharp oscillations in the Hubble parameter, $H$, and the Ricci scalar, $R$, which also have an effect. 

\begin{figure}[ht!]
\centering 
    \includegraphics[width=\textwidth]{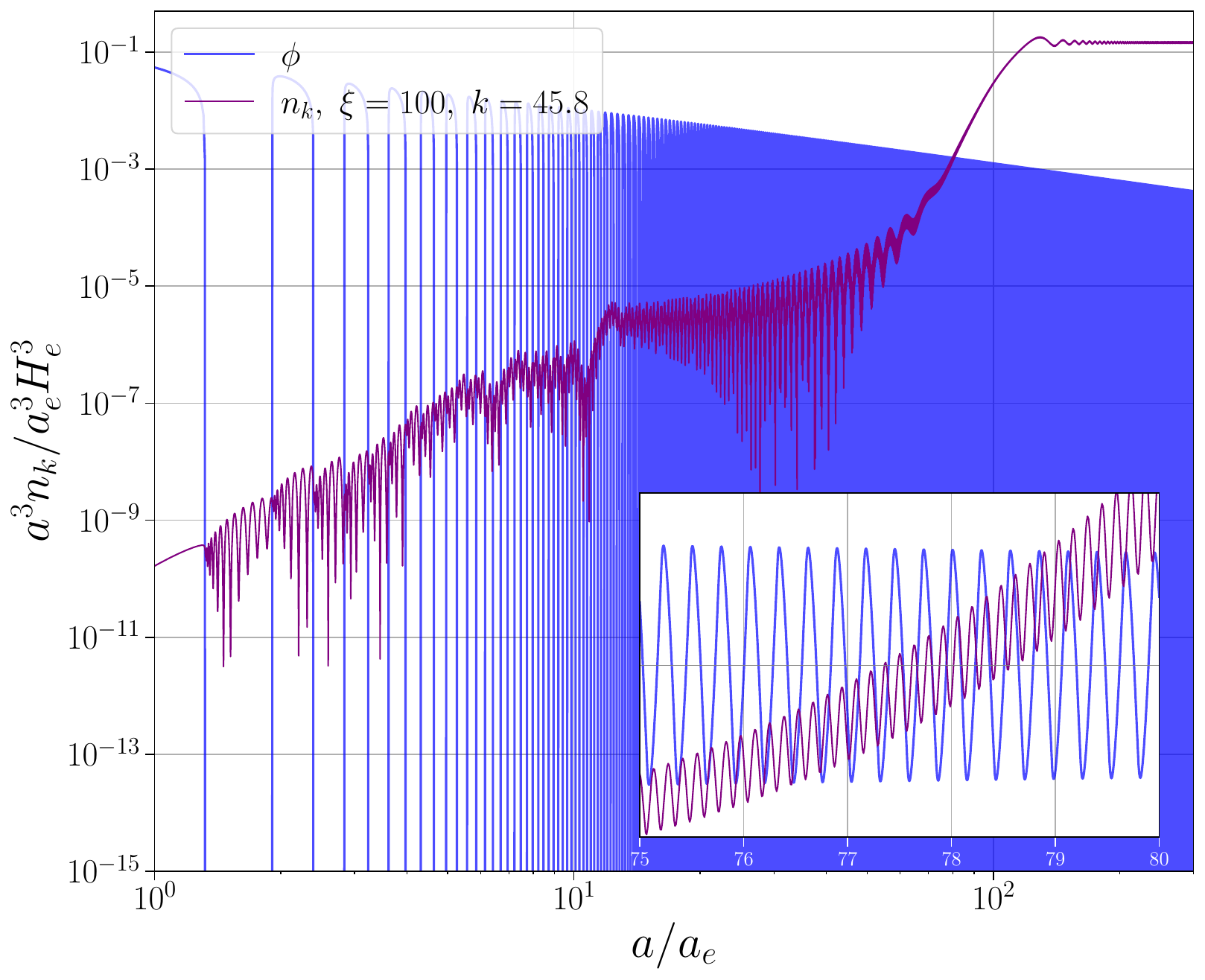}
    \caption{Time evolution of the particle number density $n_k$, for $m_\chi/H_e=0.1$, $\xi=100$, and fixed $k=45.8a_e H_e$, corresponding to the peak of the spectrum shown in Fig.~\ref{fig:spectrum_rescaled}. The particle number $n_k$ is shown in purple with $\phi$ overlaid in blue. The inset is focuses on the time $a/a_e\approx 80$ where $n_k$ shows the steepest growth, with $\phi$ and $n_k$ scaled to fit the same y-axis. This shows the rapid oscillations in $\phi$ induce the large production in $n_k$.   
} 
    \label{fig:nk_phi}
\end{figure}

Consider the $\xi=100$ scenario, which has a peak mode of $k_{\rm peak} \simeq 45.8 a_e H_e$. In Fig.~\ref{fig:nk_phi} we show the time evolution of the particle number density for this mode with respect to $a/a_e$ where $a_e$ is the scale factor at the end of inflation. The evolution of $\phi$ is superimposed to show how its oscillations induce growth in $n_k$. A zoomed-in plot shows the region in which $n_k$ grows the fastest. We can see that there are rapid oscillations in $n_k$, along with $\phi$, and that the number density reaches its asymptotic value long after the end of inflation. From this observation, we can appreciate that the significant increase in the number density is a cumulative effect arising from the background oscillations, rather than being driven by one sharp spike. In the inset, it is clear that while not exactly at the same frequency, the oscillations in $n_k$ comparable to the oscillations in $\phi$, indicating that this behavior may be driven by a resonance. 

We note that previous work on particle production in Higgs inflation has modeled the particle production from Higgs inflation as being predominantly due to early spikes in $\dot{\phi}$, and in particular, the first occurrence of the spike. The resulting production is argued to be characterized by one `spike' timescale, leading to particle production peaked on a scale $k= \xi a_e H_e$ \cite{Ema:2016dny, Babichev:2020yeo}. This approach makes the problem analytically tractable, but from Fig.~\ref{fig:nk_phi}, clearly does not accurately represent the evolution dynamics of particle production in Higgs inflation, nor the resulting late time spectrum shown in Fig.~\ref{fig:spectrum_rescaled}. We find that the spectrum of produced particles of the conformally coupled scalar is instead dominated by the cumulative effect of many rapid oscillations of the inflaton and consequent cuspy oscillations in $a(t)$, leading to features in the spectrum at $k\approx \xi^{2/3}a_e H_e$.

\begin{figure}[ht!]
\centering 
    \includegraphics[scale=0.5]{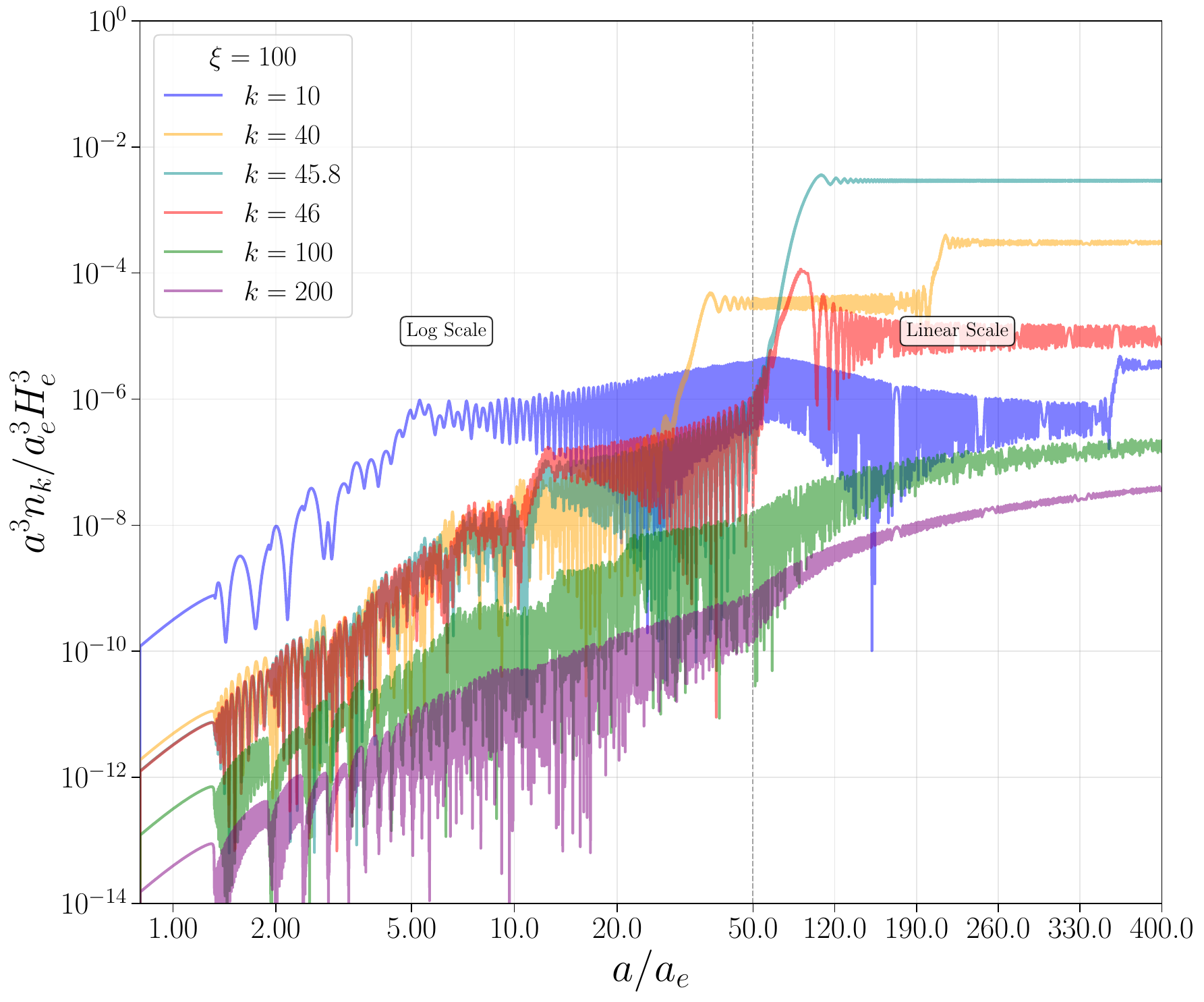}
    \caption{Multiple modes for $n_k$ are shown for $\xi=100$ and $m_{\chi}/H_e=0.1$. A resonance feature emerges for the peak mode. The x-axis is split into log and linear scales to show the evolution of particle number clearly while it grows and once it stabilizes to its final value. 
    } 
    \label{fig:nk_multiple}
\end{figure}

To see this more explicitly, it is useful to compare the behavior of the peak mode with its non-peak counterparts for this same scenario. In Fig.~\ref{fig:nk_multiple} we show the evolution of several modes, recalling that the total number density is determined at late times when the modes have reached their asymptotic value. Clearly, the peak mode dominates over the others by at least an order of magnitude in the final value. Consider the comparison of the $k=40$ (yellow), $k_{\rm peak}=45.8 a_e H_e$ (cyan), and $k=46 a_e H_e$ (red) modes. These all begin by approximately tracking each other, and then dramatically split off at some point in the evolution. This is particularly dramatic when looking at $k=46$, which is extremely near the peak but has drastically different behavior at late times and a substantially lower final number density. The sharp, rapid increase in the number density of the $k_{\rm peak}$ mode compared to its neighbors again suggests a narrow resonance feature arising from the background oscillations, analogous to preheating (see \cite{Mukhanov:2005sc} for a textbook reference). From this and Fig.~\ref{fig:nk_phi} we can appreciate that this resonance at $k \propto \xi^{2/3}$ arises from the late time behavior of $\phi$, whose frequency similarly scales as $\omega_\phi \propto \xi^{2/3}$, as discussed in Sec.~\ref{sec:Higgs} and shown in Fig.~\ref{fig:oscill_scaling}.

A reasonable concern at this point is the possibility of {\it backreaction} of the produced particles on the background dynamics.  Intensely studied in the context of preheating (for a review see e.g.~\cite{Amin:2014eta}), backreaction in this context refers to the influence of the produced particles on the background evolution. Backreaction is conventionally thought to become important once an ${\cal O}(1)$ fraction of energy density of the inflaton is transferred into particles. If backreaction becomes significant, it raises the possibility that the particle production may be modified or stopped altogether.

\begin{figure}[ht!]
\centering 
    \includegraphics[scale=0.5]{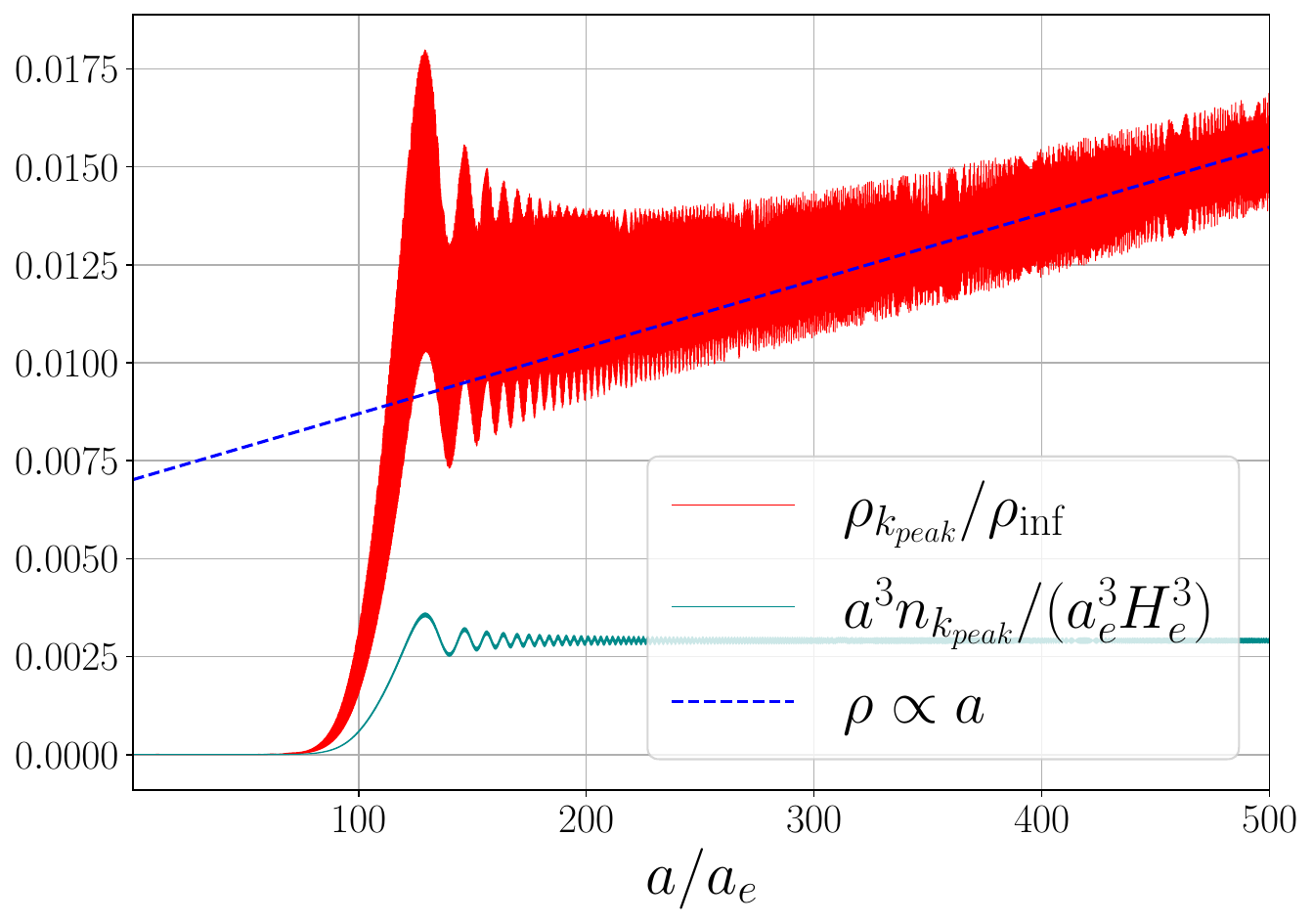}
    \caption{Relative energy density in the peak mode, for $\xi=100$ and $m_{\chi}/H_e=0.1$. Shown are the evolution of $a^3 n_k$ and the evolution of the physical energy density, $\rho_k = (\omega_k/a) n_k$, where $\omega_k ^2 = k^2+a^2 m_\chi^2$, relative to background energy density. The fractional energy density is $\ll 1$ at all times, including while the particle production is occurring. At late times, the relative energy density grows linearly in $a$, consistent with nonrelativistic particles in a radiation dominated universe. 
     } 
    \label{fig:backreaction}
\end{figure}

\begin{figure}[ht!]
\centering 
    \includegraphics[scale=0.5]{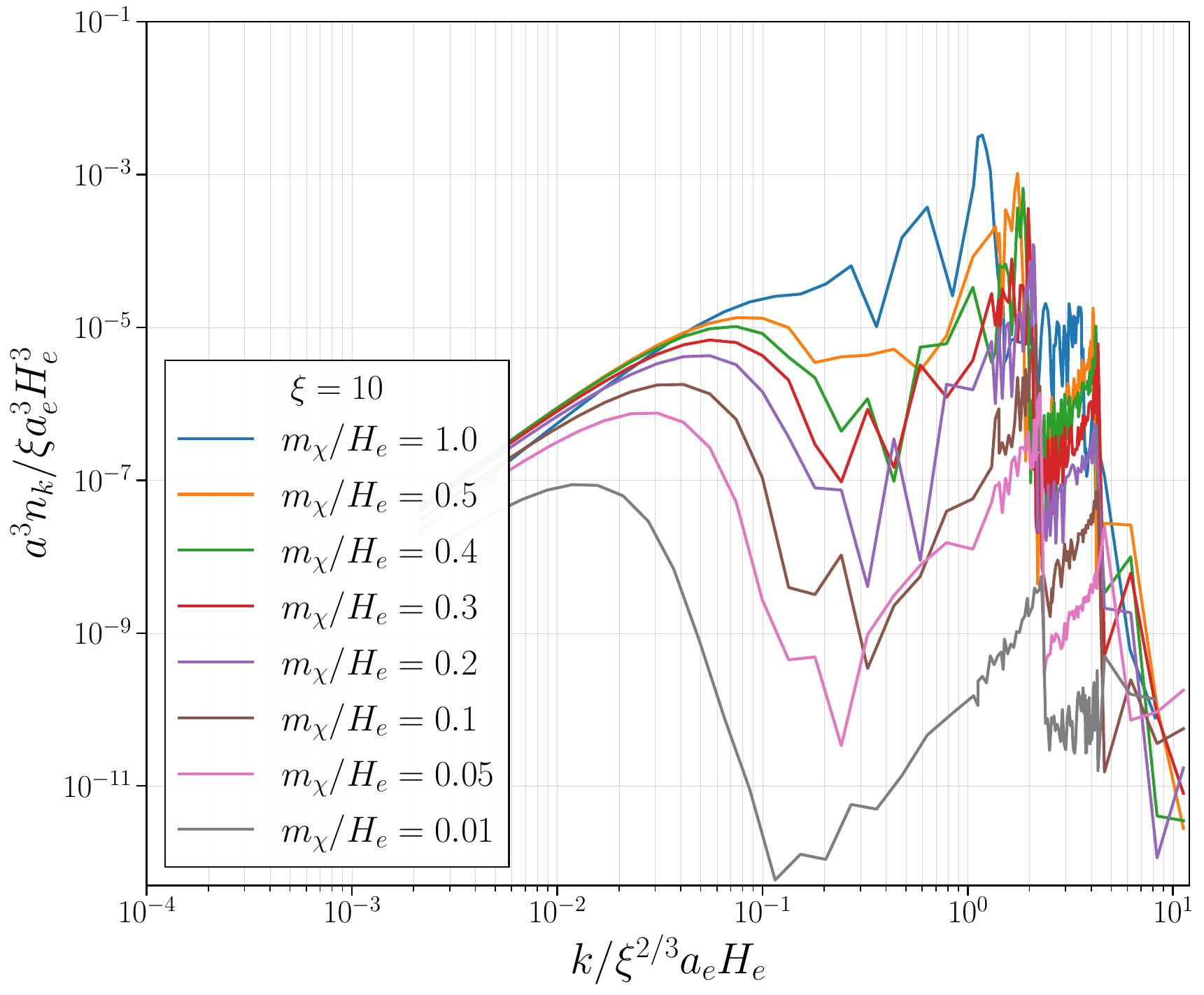}
    \caption{
    Spectrum plot for a conformally coupled scalar field with various masses ranging from $m_\chi/H_e = 0.01-1$, while keeping $\xi=10$ fixed. The particle number $n_k$ is rescaled by $\xi$ and $k$ is rescaled by $\xi^{2/3}$.    }
    \label{fig:spectrum_mass}
\end{figure}

\begin{figure}[ht!]
\centering 
    \includegraphics[scale=0.5]{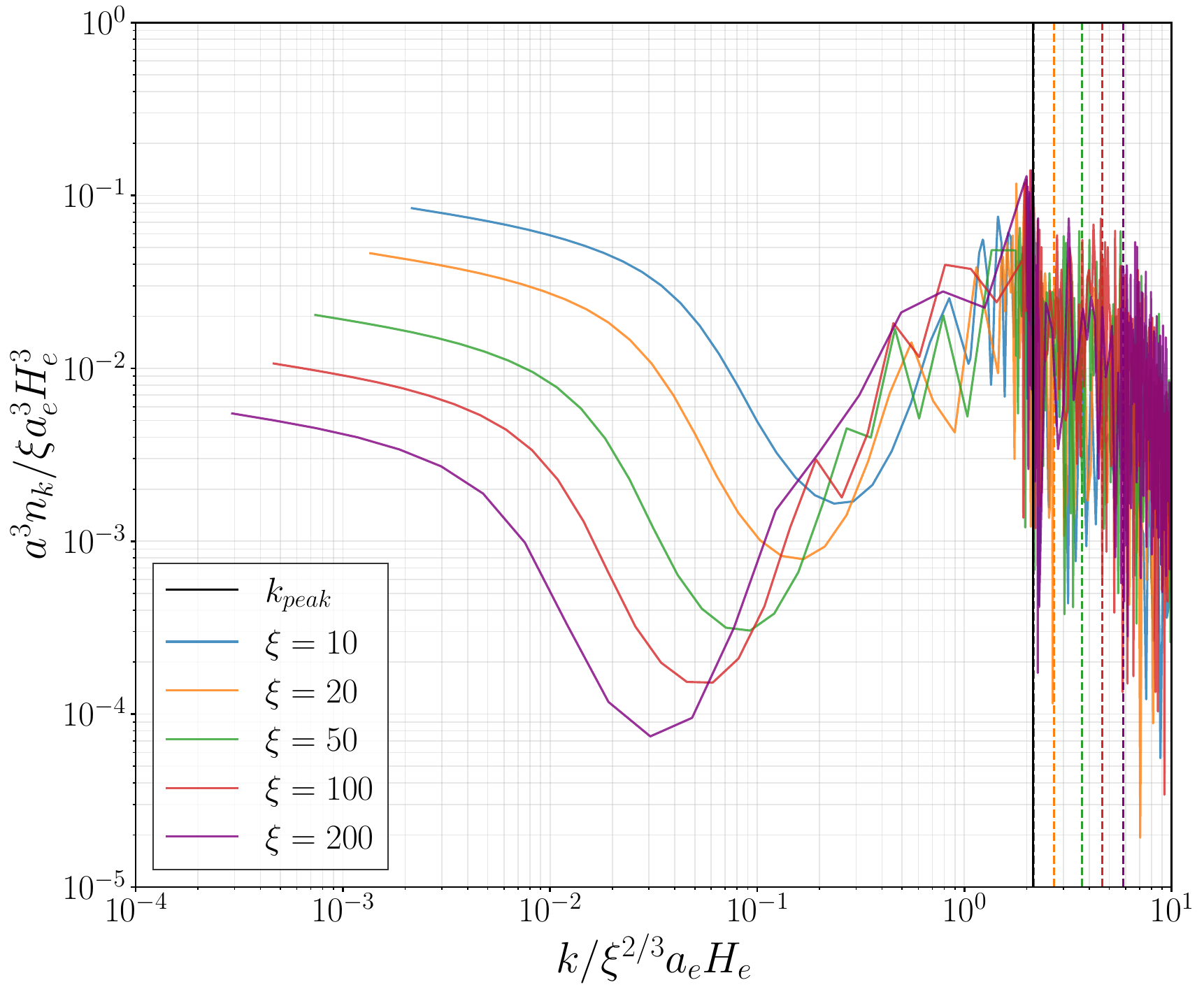}
    \caption{ 
    Particle production for a minimally coupled scalar field for various values of $\xi$ and fixed $m_{\chi}/H_e=0.1$. The particle number $n_k$ is rescaled by $\xi$ and $k$ is rescaled by $\xi^{2/3}$. Dashed lines indicate $k=\xi a_e H_e$ and we note that even in the case of minimal coupling, this wavenumber is not the peak of the spectrum. 
    } 
    \label{fig:spectrum_minimum}
\end{figure}

Here we numerically confirm that in the examples with spectra shown in Figs.~\ref{fig:spectrum_no_rescaling}, \ref{fig:spectrum_rescaled} the  produced particles comprise a negligible fraction of the energy density of the universe. In Fig.~\ref{fig:backreaction}, we compare the physical energy density in the peak mode, $\rho_{\rm peak}= \sqrt{(k/a)^2 + m_\chi^2 } n_k$, to the background energy density, namely that of the inflaton $\rho_{\rm inf}$. Since the spectrum of particles is sharply peaked, we expect the energy density to be well approximated as that of the peak mode. From Fig.~\ref{fig:backreaction} one may appreciate that $\rho_{\rm peak} / \rho_{\rm inf}\ll1$ while the particle production is occurring. At late times the fractional energy density grows linearly with the scale factor, as the particles redshift like matter  in a radiation dominated universe. Fig.~\ref{fig:backreaction} indicates that the backreaction of the CGPP on the background does not significantly alter the spectra presented here. A more careful treatment would require lattice study (see e.g.~Refs.~\cite{Nguyen:2019kbm,vandeVis:2020qcp} in the context of preheating) and a careful treatment of renormalized energy density of the quantum fluctuations (see the review Ref.~\cite{Kolb:2023ydq} for a discussion). We leave such an analysis to future work.

These results generalize to a wide range of masses. Thus far we have focused on a representative scenario with $m_{\chi}/H_e=0.1$. In Fig.~\ref{fig:spectrum_mass}, we show spectra for masses ranging from $0.01 \leq m_\chi/H_e \leq 1$ with fixed $\xi=10$. As we go to larger masses, the initial peak that also arises in quartic and quadratic inflation begins to get washed out, and the only features in the spectrum  which remain are the peaks at $k_{\rm peak}$ and $2k_{\rm peak}$. At lower masses, the peak structure remains distinct, but the contribution from $k_{\rm peak}$ and $2k_{\rm peak}$ is subdominant to the initial peak. In this case, the amplitude of the peaks depends linearly on $m_\chi$. 

Finally, we turn to the minimally coupled scalar field. The spectrum is shown in Fig.~\ref{fig:spectrum_minimum}, where we have rescaled the spectrum and $k$ as in Fig.~\ref{fig:spectrum_rescaled}. While the dynamics of this case are more complicated than those of the conformally coupled scalar because there is now an explicit factor of the Ricci scalar in the dispersion relation $\omega^2_k$,  the resulting spectrum is nonetheless peaked on the same scale as the conformally coupled scalar. This is an interesting result. From Fig.~\ref{fig:minimal-ricci} we can see that the Ricci scalar $R$ does indeed exhibit spiky, $\delta$-function-like features, generated by the background spikes in $\dot{\phi}$, of the kind modeled by Refs.~\cite{Ema:2016dny, Babichev:2020yeo}, and yet the spectrum, Fig.~\ref{fig:spectrum_minimum}, is not peaked at $k=\xi a_e H_e$. Concretely,  the spikes in $R$ reach a maximum amplitude ${\rm max}(R)= {\cal O}(1) \xi H_e ^2$ for the first spike and redshift thereafter. While ostensibly a large spike, the magnitude is such that the $a^2 R$ term in $\omega_k^2$ is subdominant to the $k^2$ term, at the time of interest, $a\sim a_e$,  for the mode of interest $k\sim\xi a_e H_e$ and $\xi \gg 1$. Therefore we expect no enhanced production of the mode $k=\xi a_e H_e$ due to the spikes in $R$. This is confirmed numerically, as we illustrate in  Fig.~\ref{fig:spectrum_rescaled}, where the mode $k=\xi a_e H_e$ is shown by dashed lines. From this one can appreciate that the production of the mode $k=\xi a_e H_e$  is always subdominant to the peak mode $k_{\rm peak} \simeq 2 \xi^{2/3} a_e H_e$.

\section{Late Time Relics}
\label{sec:latetimerelics}

\begin{figure}[htb!]
\centering 
    \includegraphics[width=0.5\textwidth]{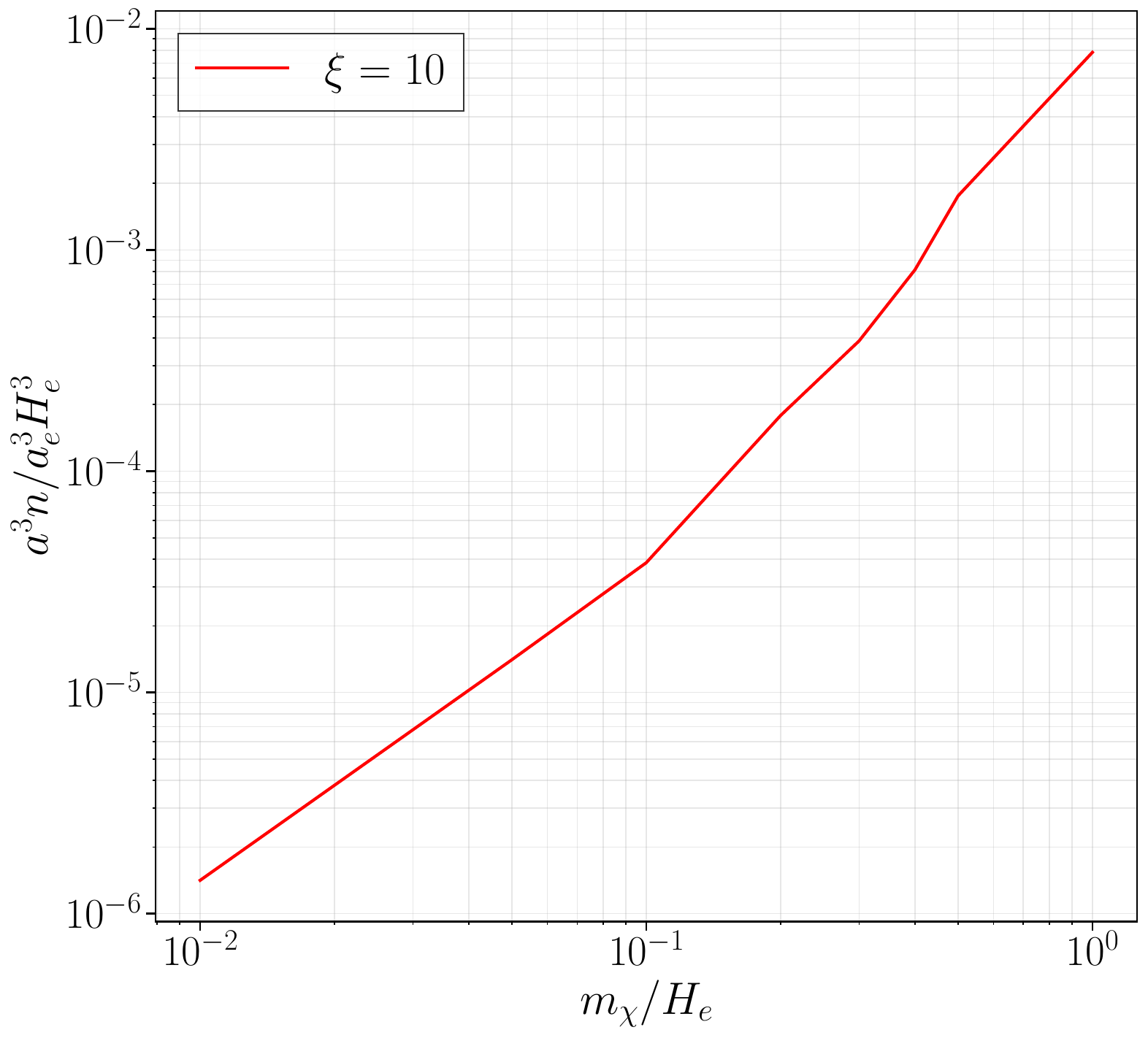}
    \includegraphics[width=0.48\textwidth]{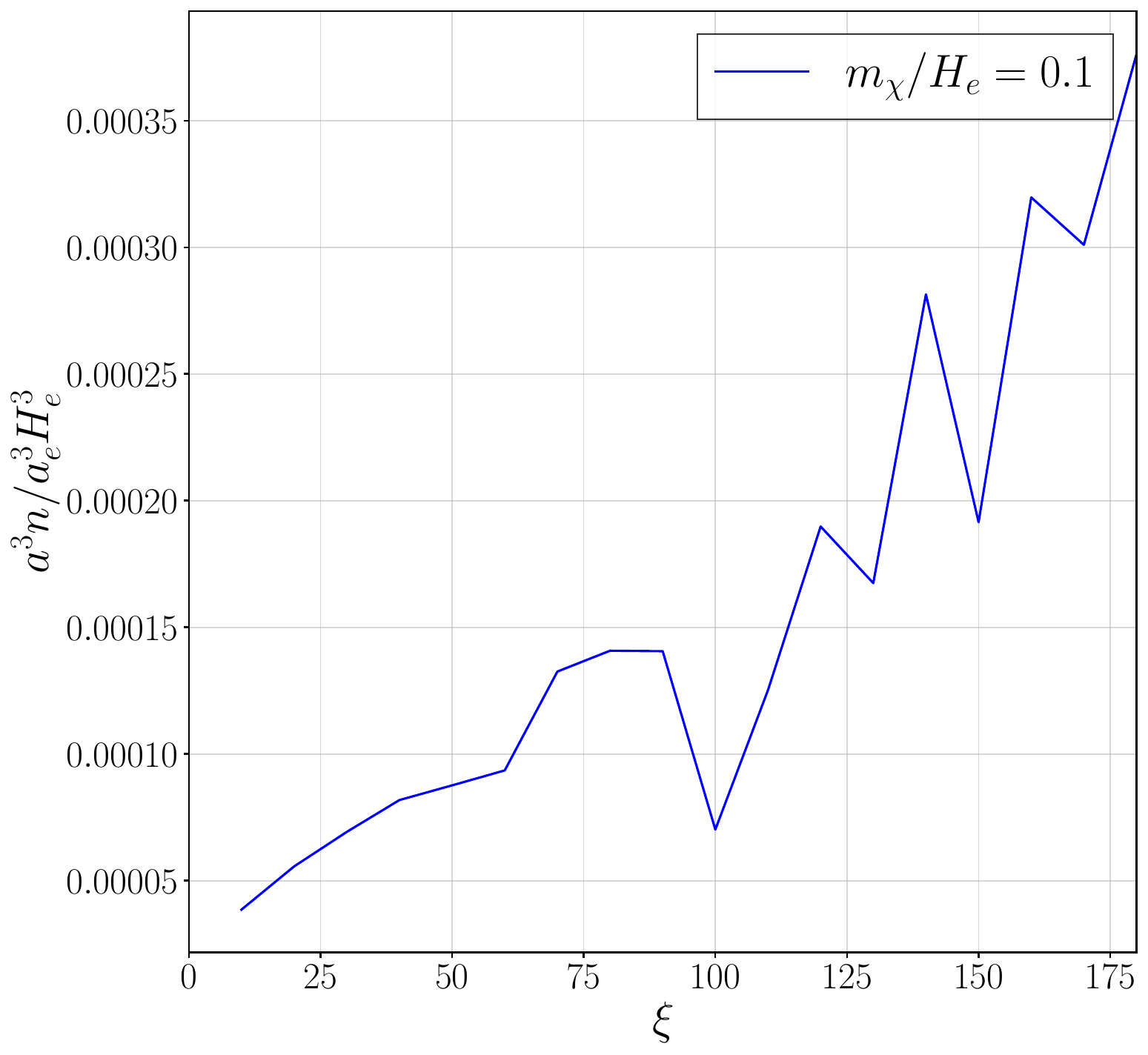}
    \caption{ Comoving number density for conformally coupled scalars, varying mass with fixed $\xi =10$ and varying $m_\chi/H_e$ (left) , and with varying $\xi$ and fixed $m_{\chi}/H_e=0.1$ (right).  } 
    \label{fig:na3_muxi}
\end{figure}

As demonstrated in the previous section, Higgs inflation exhibits an enhancement of gravitational particle production relative to minimally coupled inflation models, in the form of two new peaks in the spectrum. From the spectrum, we can obtain the comoving number density at the end of inflation, $a^3n$, by integrating over $k$, as in Eq.~\eqref{eq:na3}, then determine how this particle density can be relevant for late time relics. Two fiducial examples for $a^3n$ are shown in Fig.~\ref{fig:na3_muxi}, for varying $m_{\chi}/H_e$ at fixed $\xi=10$ (left) and varying $\xi$ at fixed $m_{\chi}/H_e=0.1$ (right). For fixed $\xi$, the number density scales approximately as $(m_\chi/H_e)^2$ in the regime of interest. Similarly, holding $m_{\chi}/H_e$ fixed and increasing $\xi$ leads to an overall increase in $n a^3$ by nearly an order of magnitude over the range $10 < \xi < 175$. 

Having characterized the number density, we can now discuss the relevant late-time phenomenology. The most natural application of the Higgs inflation particle factory is to dark matter. From the comoving number density at the end of inflation, one can determine the present-day relic density via \cite{Kolb:2023ydq}: \begin{equation}
    \frac{\Omega_{\chi}h^2 }
    {0.12}=\frac{m_{\chi}}{H_e}\left(\frac{H_e}{10^{12}\text{GeV}}\right)^2\left[\frac{T_{\rm RH}(\xi)}{10^9\text{GeV}}\right]\frac{a^3n}{10^{-5}}, \label{eq:DarkMatter}
\end{equation}
where $\Omega_\chi h^2/0.12 \sim 1 $ corresponds to the dark matter relic density today. In the above we have assumed inflation is followed by a phase of $w=0$ which transitions to $w=1/3$ at a time $t_{\rm RH}$ where $H^2(t_{\rm RH}) \propto T_{RH}^4$.  As discussed previously, the characteristic energy scale of Higgs inflation is $H_e \sim 10^{13}$ GeV, which removes $H_e$ as a free parameter. The relic density also depends on the effective reheat temperature, or more precisely the time at which the equation of state of the universe first becomes $w=1/3$, which is a $\xi$-dependent quantity, as can be appreciated from Fig.~\ref{fig:H_a_plot}. For simplicity, let us consider an instantaneous reheating scenario, which will provide an upper limit on $\Omega_\chi h^2$. In this case, $T_{\rm RH}$ is \cite{Kolb:2023ydq},  
\be 
T_{\rm RH}^{\rm inst.} \approx (8 \times 10^{14} {\rm GeV})\left(\frac{H_e}{10^{12} {\rm GeV}}\right)^{1/2}\left(\frac{g_{*\rm RH}}{106.75}\right)^{1/4},
\ee 
where $g_{*{\rm RH}}$ is the effective number of degrees of freedom in the plasma at $T_{\rm RH}$. Taking $H_e = 10^{13}$ GeV and $g_{* {\rm RH}} = 106.75$ fixes  the reheat temperature, leaving $m_{\chi}/H_e$ as the only free parameter. 

From here, there are two paths to dark matter: 
\begin{enumerate}
    \item $\chi$ itself is the dark matter,
    \item $\chi$ decays into a lighter relic, which is the dark matter.
\end{enumerate}
Let us consider both of these scenarios in turn. 

In the first case, because of the high efficiency of the particle production, there is a narrow band of parameter space where one can obtain $\Omega_\chi h^2/0.12 \sim 1$. Given the necessary high scale of inflation and $T_{\rm RH}$, for much of the parameter space $\chi$ is significantly overproduced, e.g., for $m_{\chi}/H_e=0.1$ and $\xi=100$, the relic density $\Omega_\chi h^2/0.12$ is $\mathcal{O}(10^{6})$! As a result, the region of parameter space for successful dark matter production is at lower masses such that $a^3n$ is sufficiently suppressed, near, e.g., $m_{\chi} \sim 10^{10}$ GeV for $\xi=10$. In this scenario, the second peak in the spectrum at $k_{\rm peak}$ is suppressed, and the particle production is dominated by the first peak, as in quadratic and quartic inflation models (see \cite{Kolb:2023ydq}). Therefore, it is possible to realize a dark matter scenario in windows of the $\{m_{\chi}/H_e, \xi\}$ parameter space such that $a^3n$ is not overly enhanced. Additionally, considering later reheating can also help to widen the parameter space.

On the other hand in regions where the production is significantly enhanced, we can consider the scenario in which the gravitationally produced scalars decay into lighter particles, which can then be the dark matter. We assume that the entire density of $\chi$ particles decay into a stable dark matter candidate, $\chi'$, via the process
\be 
\chi \rightarrow \chi'\chi'.
\ee 
This implies that $n_{\chi}' = 2 n_{\chi}$. If we further assume the $\chi'$ particles are non-relativistic at late times, then relic density of $\chi'$ can be determined from Eq.~\ref{eq:DarkMatter} as 
\be 
\frac{\Omega_{\chi'} h^2}{0.12} = 2\frac{m_{\chi'}}{m_\chi}\frac{\Omega_\chi h^2}{0.12},
\ee 
where $m_{\chi'}$ is the dark matter mass and $\Omega_\chi h^2$, defined by Eq.~\ref{eq:DarkMatter}, is the relic density that would be in $\chi$ were it not for the decay into $\chi'$. 

\begin{figure}[htb!]
\centering 
    \includegraphics[width=0.49\textwidth]{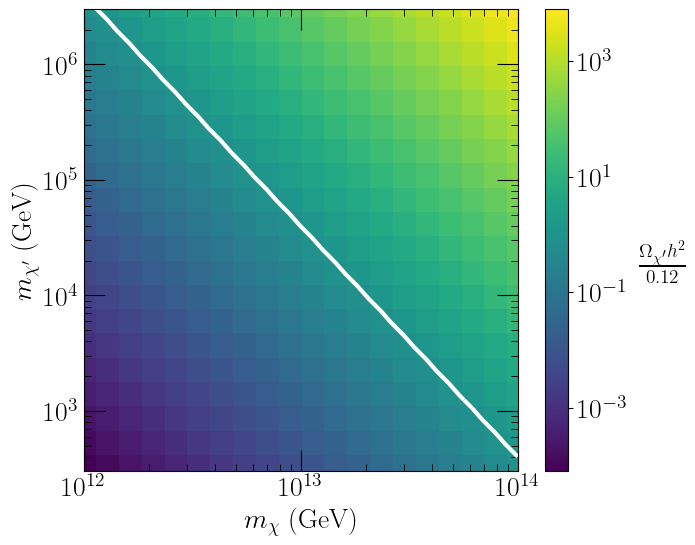}    \includegraphics[width=0.49\textwidth]{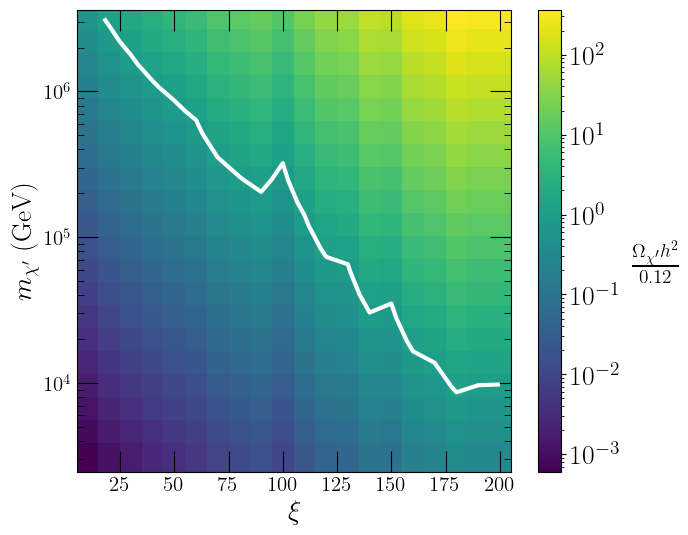}
    \caption{
    Parameter space for the decay product of the gravitationally produced scalars to account for the dark matter relic density at fixed $\xi=100$ (left) and fixed $m_\chi/H_e = 0.1$. The white lines correspond to $\Omega_{\chi'}h^2/0.12 = 1$. 
    }
    \label{fig:DMdecay}
\end{figure}

This allows us to appreciably widen the parameter space where $\chi'$ can account for the dark matter. \Cref{fig:DMdecay} shows $\Omega_{\chi'} h^2$ in the plane of $\{ m_\chi, m_{\chi'}\}$ for fixed $\xi=10$. For example, for $m_{\chi}/H_e = 0.1$, the necessary decay mass $m_{\chi'}$ to yield $\Omega_{\chi'}h^2=0.12$ is $m_{\chi'} \sim 10^7 $ GeV. Similarly, for $\xi=150$ and fixed $m_\chi/H_e=0.1$, the decay product mass needs to be $\mathcal{O}$(10 TeV). Interestingly, these decay products could be well within the reach of terrestrial colliders as well as dark matter direct detection experiments, providing a potential new window into the early universe via these distinctive late-time relics. 

\section{Discussion}
\label{sec:Discussion}

In this work we have studied the gravitational particle production in a Higgs inflation scenario, modeling the inflaton as a scalar field with a nonminimal coupling to gravity, $\xi$, which can be realized as the Higgs of the Standard Model, but does not necessarily have to be. To this end, we have considered a range of $\xi$ away from the usual high values of $\xi\sim 10^4$. For the inflationary background dynamics, we found universal scaling relations for the slow-roll parameter $\varepsilon$ and the Ricci scalar $R$ during the post-inflationary oscillations, which scale linearly with $\xi$, and a universal scaling relation for the asymptotic oscillation frequency of the Higgs, which scales like a fractional power as $\xi^{2/3}$. 

These features of the background contribute to a striking universality in the spectrum of particle production for scalar spectator fields, both conformally and minimally coupled, in particular the emergence of new peaks which scale as $k\propto \xi^{2/3}a_eH_e$. From the large features of the spectrum, we found that Higgs inflation provides a highly efficient particle factory, producing a high number density of particles, which we showed could easily be the dark matter, or decay into the dark matter for a wide range of our $\{m_\chi, \xi\}$ parameter space which is potentially accessible with terrestrial colliders and/or dark matter direct detection experiments. This motivates a dedicated scan of parameter space for dark matter in Higgs inflation, beyond the preliminary explorations we have performed here.

It is interesting to contemplate other phenomenological applications of this particle factory. One particularly intriguing direction is gravitational reheating \cite{Bassett:1997az,Haque:2022kez}, in which the reheating of the universe occurs via fields which are only gravitationally coupled. Another intriguing possibility is that the gravitationally produced particles could undergo out-of-equilibrium decay to generate the observed baryon asymmetry of the universe, similar to  \cite{Hashiba:2019mzm,Bernal:2021kaj,Co:2022bgh,Barman:2022qgt,Fujikura:2022udt,Flores:2024lzv}. Finally, another potential observational signature of the particle factory is the scalar-induced gravitational waves \cite{Assadullahi:2009jc,Inomata:2018epa,Domenech:2021ztg} generated by the sharp spikes in the particle spectrum of CGPP in Higgs inflation. It is possible that such gravitational waves could be accessible with future detectors and provide yet another window into these early universe dynamics.

On the inflation side, an interesting question is whether the results presented here generalize to other models of inflation that feature a nonminimal coupling to gravity, such as Higgs-$R^2$ inflation \cite{He:2018gyf}, variations on natural inflation \cite{Lorenzoni:2024krn,McDonough:2020gmn}, models that can seed primordial black holes \cite{Geller:2022nkr,Qin:2023lgo}, and in models of inflation motivated by recent data (for an overview see \cite{Kallosh:2025ijd,Ferreira:2025lrd} and references therein). We leave this and other interesting directions to future work.

\acknowledgments
We thank David Kaiser, Andrew Long, and Sarunas Verner, for useful comments and discussions. LJ is supported by the Provost's Postdoctoral Fellowship at Johns Hopkins University and was supported by the Kavli Institute for Cosmological Physics at the University of Chicago while this work was carried out. EWK is supported in part by the Kavli Institute for Cosmological Physics at the University of Chicago. TC and EM are supported in part by the Arthur B. McDonald Canadian Astroparticle Physics Institute and by the Natural Sciences and Engineering Research Council of Canada.

\appendix

\clearpage

\section{Metric Signature and Sign Conventions}
\label{app:sign_conventions}

In this appendix, we provide an overview of the sign conventions of the inflationary background evolution equations. 

\subsection{Background Evolution}

The action for a scalar field coupled to gravity  is given by
\begin{align}
    S[\phi(\eta,x)]=\int d\eta \int d^3x~\sqrt{-g} \,\cal{L}
\end{align}
where ${\cal{L}}$ is given by \cite{Kolb:2023ydq}
\begin{align}
    {\cal{L}}=-s_1\frac{1}{2}g^{\mu\nu}\partial_{\mu}\phi\partial_{\nu}\phi-V(\phi)-s_1s_3\frac{1}{2}\xi R\phi^2,
\end{align}
where $s_1$ defines the metric signature,
\begin{equation}
    g_{\mu \nu} ^{\rm MINK} = \eta_{\mu \nu} = s_1 {
    \rm diag
    }(-1,+1,+1,+1),
\end{equation}
and $s_3$ defines the sign of Einstein-Hilbert action,
\begin{equation}
    S_{\rm EH} = s_1 s_3 \frac{1}{2}M_\mathrm{Pl}^2 \int d^4 x \sqrt{-g}R
\end{equation}
Using the metric $g_{\mu\nu}^{FLRW}(x)=s_1a^2(\eta)~\rm diag(-1,+1,+1,+1)$, $g^{\mu\nu}=s_1\frac{1}{a^2}\eta^{\mu\nu}$ and the determinant $\sqrt{-g}=a^4(\eta)$, we find
\begin{align}
    S[\phi(\eta,x)]=\int d\eta \int d^3x~\left[\left(\frac{1}{2}a^2(\partial_{\eta}\phi)^2-\frac{1}{2}a^2(\nabla\phi)^2\right)-a^4V(\phi)-s_1s_3\frac{1}{2}a^4\xi R\phi^2\right]
\end{align}
where we used $s_1^2=1$ in the kinetic term.

Now we are in a position to derive the equations of motion. To do so we will make use of the scalar sector of the ADM formalism. Namely we decompose the metric as
\begin{equation}
    g_{\mu \nu}= s_1 {\rm diag}( - N^2(t) , a^2(t), a^2(t) ,a^2(t)).
\end{equation}
We start from the action,
\begin{equation}
    S = \int d^4x \sqrt{-g}\left[ \frac{1}{2}s_1 s_3(M_\mathrm{Pl}^2 + \xi \phi^2)R -s_1\frac{1}{2}g^{\mu\nu}\partial_{\mu}\phi\partial_{\nu}\phi-V(\phi)\right]
\end{equation}
and assume $\phi=\phi(t)$.

To get the field equations we insert the metric and background field into the action. Note the Ricci scalar is given by
\begin{equation}
    R = \frac{6 N(t) \left(a(t) \ddot{a}(t)+\dot{a}^2(t)\right)-6 a(t) \dot{a}(t) \dot{N}(t)}{s_1 s_3 a^2(t) N^3(t)}
\end{equation}
In total the Lagrangian is given by,
\begin{eqnarray}
    {\cal L} = && \frac{3 M_\mathrm{Pl}^2 s_1^2 a^2(t) \ddot{a}(t)}{N(t)} + \frac{3 \xi  s_1^2 a^2(t) \phi^2 (t) \ddot{a}(t)}{N(t)}\nonumber \\
    && -\frac{3 M_\mathrm{Pl}^2 s_1^2 a^2(t) \dot{a}(t) \dot{N}(t)}{N^2(t)}+\frac{3 M_\mathrm{Pl}^2 s_1^2 a(t) \dot{a}^2(t)}{N(t)}-\frac{3 \xi  s_1^2 a^2(t) \phi^2(t) \dot{a}(t) \dot{N}(t)}{N^2(t)} \nonumber \\
    &&+\frac{3 \xi  s_1^2 a(t) \phi^2(t) \dot{a}^2(t)}{N(t)}-s_1^2 a^3(t) N(t) V(\phi (t))+\frac{s_1^2 a^3(t) \dot{\phi}^2(t)}{2 N(t)}.
\end{eqnarray}
Note that $s_3$ has dropped out entirely and $s_1$ only appears as $s_1^2=1$. Thus the Lagrangian and hence equations of motion are independent of $s_1$ and $s_3$.

To derive the equations of motion, we vary the Lagrangian with respect to the fields to find the Euler-Lagrange equations. We then fix $N(t)=1$ in the equations of motion to restrict to FRW solutions. Varying with respect to $\phi$, and setting $N=1$, we find
\begin{equation}
    \ddot{\phi}+ 3 H \dot{\phi} - 6  \xi \phi \left( H^2 + \dot{H} \right) + \frac{dV}{d\phi}=0
\end{equation}
where we used $H=\dot{a}/a$. Varying with respect to $N(t)$, we find
\begin{equation}
     3   M_\mathrm{Pl}^2 H^2 + \xi \phi^2 \left( H^2 + \frac{d \log \phi}{dt}\right) = \frac{1}{2}\dot{\phi^2} + V(\phi).
\end{equation}
We note these equations are independent of the sign convention.

\subsection{Mode Equations for Spectator field CGPP }

Let's now shift gears and consider the mode equations for the CGPP of a spectator field, $\varphi$. To canonically normalize the field, we perform the following field redefinition,
\begin{align}
    &\varphi(\eta,x)=\frac{1}{a(\eta)}\chi(\eta,x), \nonumber \\
    &\partial_{\eta}(\varphi(\eta,x))=-\chi H+\frac{1}{a}\partial_{\eta}\chi
\end{align}
so that the action becomes
\begin{align}
    S[\varphi(\eta,x)]&=\int d\eta \int d^3x~\left[\frac{1}{2}a^2\left(-\chi H+\frac{1}{a}\partial_{\eta}\chi\right)^2-\frac{1}{2}(\nabla\chi)^2-\frac{1}{2}a^2m^2\chi^2-\frac{s_1 s_3}{2}a^2\xi R\chi^2\right] \nonumber \\
    &=\int d\eta \int d^3x~\left[\frac{1}{2}(\partial_{\eta}\chi)^2-aH\chi\partial_{\eta}\chi-\frac{1}{2}(\nabla\chi)^2-\frac{1}{2}a^2\chi^2(m^2+s_1 s_3\xi R-H^2)\right] .
\end{align}
The term $aH\chi\partial_{\eta}\chi$ can be rewritten by noting 
\begin{align}
\label{eq:term}
    -\frac{1}{2}\partial_{\eta}(aH\chi^2)&=-\frac{1}{2}a'H\chi^2-\frac{1}{2}a\partial_{\eta}\left(\frac{a'}{a^2}\right)\chi^2-aH\chi\partial_{\eta}\chi \nonumber\\
    &=-\frac{1}{2}a'H\chi^2-\frac{1}{2}\chi^2\frac{a''}{a} +\chi^2\frac{a'^2}{a^2}-aH\chi\partial_{\eta}\chi .
\end{align}
We note we can express $a''/a$ in terms of the Ricci scalar as (\cite{Kolb:2023ydq}):  
\begin{equation}
    \frac{a''}{a}=\frac{1}{6}s_1 s_3 a^2 R
\end{equation}
and we note $H=\frac{a'}{a^2}$.
We rearrange Eq.~\eqref{eq:term} for $a H \chi \partial_\eta \chi$ to find
\begin{align}
    aH\chi\partial_{\eta}\chi &= \frac{1}{2}\partial_{\eta}(aH\chi^2)-\frac{1}{2}a'H\chi^2-\frac{1}{2}\chi^2\frac{a''}{a}+\chi^2a'H \nonumber \\
    &=\frac{1}{2}\partial_{\eta}(aH\chi^2)+\frac{1}{2}H^2a^2\chi^2-\frac{1}{2}\chi^2\left(\frac{1}{6}s_1s_3a^2R\right) \nonumber\\
    &=\frac{1}{2}\partial_{\eta}(aH\chi^2)+\frac{1}{2}a^2\left(H^2-\frac{1}{6}s_1s_3R\right)\chi^2.
\end{align}
We put this back into action,
\begin{align}
    S[\varphi(\eta,x)]=&\int d\eta\int d^3x\left[\right.\frac{1}{2}(\partial_{\eta}\chi)^2-\frac{1}{2}\partial_{\eta}(aH\chi^2)-\frac{1}{2}a^2\left(H^2-\frac{1}{6}s_1s_3R\right)\chi^2-\frac{1}{2}(\nabla\chi)^2 \nonumber\\
    &-\frac{1}{2}a^2\chi^2(m^2+s_1s_3\xi R-H^2)\left. \right] \nonumber\\
    =&\int d\eta\int d^3x\left[\frac{1}{2}(\partial_{\eta}\chi)^2-\frac{1}{2}(\nabla\chi)^2-\frac{1}{2}a^2\chi^2(m^2-s_1s_3\left(\frac{1}{6}-\xi \right)R)\right] \label{eqn:exp-action},
\end{align}
where we have dropped the total derivative term $-\frac{1}{2}\partial_{\eta}(aH\chi^2)$ to obtain
\begin{equation}
    aH\chi\partial_{\eta}\chi = \frac{1}{2}\partial_{\eta}(aH\chi^2)+\frac{1}{2}a^2(\frac{-s_1 s_3}{6}R+ H^2)\chi^2.
\end{equation}
We plug this into the action, and drop the total derivative term $\partial_{\eta}(aH\chi^2)$  to arrive at
\begin{equation}
\label{eq:Sfinal}
    S[\varphi(\eta,x)]=\int d\eta \int d^3x~\left[\frac{1}{2}(\partial_{\eta}\chi)^2-\frac{1}{2}(\nabla\chi)^2-\frac{1}{2}a^2 m _{\rm eff}^2 \chi^2\right],
\end{equation}
with 
\begin{equation}
    m_{\rm eff}^2 = m^2 - s_1 s_3\left( \frac{1}{6}-  \xi\right)R
\end{equation}
Note that the first two terms in eq.~\eqref{eq:Sfinal} are independent of $s_1$: they inherited an $s_1 ^2  = 1$.

Finally the mode equation for Fourier modes $\chi_k$ is
\begin{equation}
    \chi_k '' + \omega_k ^2 \chi_k =0
\end{equation}
with 
\begin{equation}
    \omega_k ^2 = k^2 + a^2 m^2 -  s_1 s_3\left( \frac{1}{6}-  \xi\right)a^2R
\end{equation}
which generalizes Eq.~12 of \cite{Kolb:2023ydq} to the case of general $s_1$, $s_3$.

\bibliographystyle{JHEP} 
\bibliography{ref}

\end{document}